\documentclass[twocolumn]{aastex61}

\usepackage[maxfloats=40]{morefloats} 
\usepackage{color}
\usepackage{catoptions}
\usepackage{booktabs}
\usepackage{amsmath}
\usepackage[all]{hypcap} 
\newcommand{\Rom}[1]{\uppercase\expandafter{\romannumeral #1\relax}}

\begin{document}
\makeatletter

\renewcommand{\sectionautorefname}{Section\negthinspace}
\renewcommand{\subsectionautorefname}{Section\negthinspace}
\renewcommand{\subsubsectionautorefname}{Section\negthinspace}

\title{MULTI-WAVELENGTH STELLAR POLARIMETRY OF THE FILAMENTARY CLOUD IC5146: \Rom{1}. DUST PROPERTIES}
\maxdeadcycles=1000

\author{Jia-Wei Wang}
\affil{Institute of Astronomy , National Tsing Hua University, 101, Section 2, Kuang-Fu Road, Hsinchu 30013, Taiwan, R.O.C}
\email{jwwang@gapp.nthu.edu.tw}

\author{Shih-Ping Lai}
\affil{Institute of Astronomy \& Department of Physics, National Tsing Hua University, 101, Section 2, Kuang-Fu Road, Hsinchu 30013, Taiwan, R.O.C}
\email{slai@phys.nthu.edu.tw}

\author{Chakali Eswaraiah}
\affil{Institute of Astronomy \& Department of Physics, National Tsing Hua University, 101, Section 2, Kuang-Fu Road, Hsinchu 30013, Taiwan, R.O.C}
\affil{Institute of Astronomy, National Central University, 300 Jhongda Rd, Taoyuan 32001, Taiwan, R.O.C}
\affil{Aryabhatta Research Institute of Observational Sciences (ARIES), Manora-peak, Nainital, Uttarakhand-state, 263002, India}
\email{eswarbramha@gmail.com}

\author{Dan P. Clemens}
\affiliation{Institute for Astrophysical Research, Boston University, 725 Commonwealth Avenue, Boston, MA 02215, USA}
\email{clemens@bu.edu}

\author{Wen-Ping Chen}
\affil{Institute of Astronomy, National Central University, 300 Jhongda Rd, Taoyuan 32001, Taiwan, R.O.C}
\email{wchen@gm.astro.ncu.edu.tw}

\author{Anil K. Pandey}
\affil{Aryabhatta Research Institute of Observational Sciences (ARIES), Manora-peak, Nainital, Uttarakhand-state, 263002, India}
\email{pandey@aries.res.in}

\begin{abstract}
We present optical and near-infrared stellar polarization observations toward the dark filamentary clouds associated with IC5146.
The data allow us to investigate the dust properties (this paper) and the magnetic field structure (Paper \Rom{2}). 
A total of 2022 background stars were detected in $R_{c}$-, $i'$-, $H$-, and/or $K$-bands to $A_V \lesssim 25$ mag. The ratio of the polarization percentage at different wavelengths provides an estimate of $\lambda_{max}$, the wavelength of peak polarization, which is an indicator of the small-size cutoff of the grain size distribution. 
The grain size distribution seems to significantly change at $A_V \sim$ 3 mag, where both the average and dispersion of $P_{R_c}/P_{H}$ decrease. 
In addition, we found $\lambda_{max}$ $\sim$ 0.6--0.9 $\mu$m for $A_V>2.5$ mag, which is larger than the $\sim$ 0.55 $\mu$m in the general ISM, suggesting that grain growth has already started in low $A_V$ regions. Our data also reveal that polarization efficiency (PE $\equiv P_{\lambda}/A_V$) decreases with $A_V$ as a power-law in $R_c$-, $i'$-, and $K$-bands with indices of -0.71$\pm$0.10,  -1.23$\pm$0.10 and -0.53$\pm$0.09. However, $H$-band data show a power index change; the PE varies with $A_V$ steeply (index of -0.95$\pm$0.30) when $A_V < 2.88\pm0.67$ mag but softly (index of -0.25$\pm$0.06) for greater $A_V$ values. The soft decay of PE in high $A_V$ regions is consistent with the Radiative Aligned Torque model, suggesting that our data trace the magnetic field to $A_V \sim 20$ mag. Furthermore, the breakpoint found in $H$-band is similar to the $A_V$ where we found the $P_{R_c}/P_{H}$ dispersion significantly decreased. Therefore, the flat PE-$A_V$ in high $A_V$ regions implies that the power index changes result from additional grain growth. 
\end{abstract}


\section{INTRODUCTION}
\label{sec:intro}
The linearly polarized background starlight observed in early studies \citep{ha49,hi49}, was suggested to result from dichroic absorption by non-spherical dust grains aligned to B-fields \citep{hi49b}. Hence, polarized starlight is commonly used as a tracer of the B-field structure in the plane of sky \citep[e.g.,][]{ch11,cl12a,es12}. 
The first theory to explain dust alignment utilized a paramagnetic mechanism (DG alignment) \citep{da51}.
However, later studies revealed time-scale and efficiency problems, so other alignment mechanisms were offered, including superparamagnetic grains \citep{jo67} and superthermal rotation rates \citep{pu79}. But most of these are efficient only for particular physical conditions and fail to explain the polarization observed across the wide variety of environments (see review in \citealt[]{an15}).

Radiative Aligned Torques (RATs) \citep{do76,dr96,dr97,la07} is currently the best model to explain how dust grains align with B-fields. The RATs theory assumes an anisotropic radiation field impacting non-spherical dust grains. The radiation field can generate net torque on the grains and induce both spin of the grains and precession about the B-field. For typical interstellar radiation fields, the RATs alignment timescale is much faster than DG alignment \citep{la07}, and superthermal rotation tends to enhance RATs alignment more than DG alignment \citep{ho09}. In addition, RATs is efficient across a variety of environments, and thus more likely dominating the grain alignment process.

A key prediction of RATs theory is the decreasing polarization efficiency (PE, P$_{\lambda}$/$A_V$) with increasing $A_V$, because the radiation field required to align the dust grains is decreased due to extinction \citep{la97}. In order to examine whether the RATs theory can explain observed polarizations, 
\citet{wh08} performed a numerical simulation, based on RATs theory, and assuming a starless core with only  an external radiation field and a fixed dust grain size distribution for all $A_V$. Their simulation results 
matched their $K$-band polarimetry data toward Taurus up to at least $A_V \sim 10$ mag; the $K$-band polarization degree varied with $A_V$ as a power-law, with $P_\lambda/\tau_K \propto A_V^{-0.52}$. 
In addition, their simulation predicted the RATs mechanism would cease as $A_V$ approached 10 mag, 
since the radiation that can penetrate to such high column densities has wavelengths too long to effectively align the grains. However, the single power-law well matched their data over the $A_V=$ 0--30 mag range, which was inconsistent with the predicted cessation of alignment at $A_V \sim$ 10 mag. To explain the inconsistency, \citet{wh08} argued that the RAT mechanism may still work in high $A_V$ regions if: (1) the dust grains have undergone significant growth, and so can couple to longer wavelength radiation, or; 
(2) embedded stars are present to provide radiation at shorter wavelength to enhance the alignment of small dust grains.

To further test the RATs theory in dense clouds, \citet{jo15} used both infrared and submillimeter polarimetry data to trace the variation of polarization efficiency up to $A_V \sim 100$ mag within starless cores. They found a change of power-law index for polarization efficiency versus extinction from -0.5 to -1 at $A_V \sim 20$ mag, consistent with the cessation of alignment predicted by \citet{wh08}. In addition, \citet{jo16} found a deeper break point at $A_V \sim$ hundreds mag and a reversed trend such that the power-law index of polarization efficiency versus $A_V$ changed from -1 to -0.5 within the Class 0 YSO G034.43+00.24 MM1. The opposite trends shown in these two cases suggest that dust alignment in dense clouds could be strongly affected by environmental effects.

In the RATs paradigm, the polarization degree is expected to be wavelength-dependent, 
because the cloud extinction penetration is wavelength-dependent and 
because the highest alignment efficiency occurs
at wavelengths similar to the sizes of the grains \citep{ch05,la07}. 
Therefore, the polarization wavelength-dependence is determined by both the extinction and the size of dust grains,
resulting in a polarization spectrum exhibiting a single peak ($\lambda_{max}$).
\citet{ki95} showed that $\lambda_{max}$ is sensitive to the small-size cutoff of the grain size distribution,
since the dust size distribution is a power-law and the small-size cutoff will effectively determine the mean size of the dust grains. \citet{wh08} calculated the $\lambda_{max}$ variation with $A_V$ using a RATs model with a fixed grain size distribution over $A_V$, and showed that the RATs theory can explain their observed dependence of $\lambda_{max}$ increasing with $A_V$ to $\sim$ 6\,mag, but that RATs predicts $\lambda_{max}$ should flatten for larger $A_V$ in the Taurus cloud.
 
Recent observational results are mostly consistent with RATs theory (\citealt{wh08,al14,ca14}), although the dust alignment efficiency varies from source to source. 
As an extreme case, \citet{go95} found a constant near-infrared polarization degree toward L1755 for $A_V$ of 1--10 mag, and argued that near-infrared polarization can only trace the B-field on the surfaces of dark clouds. In contrast, other observations mostly showed polarization degree increasing with $A_V$, such as $P\propto A_V^{0.48}$ toward Taurus \citep{wh08} and $P\propto A_V^{0.26}$ toward L204 cloud 3 \citep{ca14}, suggesting that the constant polarization degree found in \citet{go95} represented a special case and was not the norm. 

It is still unclear why the P-$A_V$ relation found in L1755 is different from the relation found for other clouds. \citet{wh08} re-analyzed the L1755 polarimetry data with better background star selection and extinction estimation, but still found the same relation. They speculated that this unique relation might result from a lack of the typical physical conditions leading to grain alignment inside this cloud or a reduction of observed polarization due to complex B-field structure within the L1755 filaments. To examine these possibilities, more polarimetry observations covering both a wider area and more wavebands could help to reveal the alignment conditions for dust grains with different sizes and to probe the detailed B-field structure. In addition, since the \citet{go95} polarization detections only spanned a small range of $A_V$ (2--8 mag), it would be of interest to perform polarization observations with higher sensitivity to investigate this relation to higher $A_V$ values.

An ideal target for detailed testing the RATs paradigm is the system of dark cloud filaments associated with the reflection and emission nebula IC5146, a nearby ($\sim$ 460 pc) star-forming region in Cygnus. The system consists of a young cluster inside an H\Rom{2} region and a long, main, dark cloud filament with several sub-filaments branching out from the main filament. 
The \textit{Herschel} Gould Belt Survey \citep{ar11} revealed a complex network of filaments within the long dark cloud highlighting the locations of young forming stars. The \textit{Planck} polarization map of IC5146 \citep{pl15} revealed that the B-field is nearly uniform, with the mm-wavelength-traced B-field oriented perpendicular to the elongation orientation of the main filament. Such an apparently uniform B-field gives the IC5146 dark cloud system an advantage for testing the RATs model, because the high angular resolution of stellar polarizations can be used to test for, and quantify, complex B-fields along the multiple line of sights.

In a series of papers, we will report measurements and analyses of the polarizations of stars behind the IC5146 filamentary cloud. The polarimetry of background starlight was performed at both optical and infrared wavelengths to probe the large-scale B-field in the cloud. In this paper, we focus on the dust properties and the dust alignment conditions revealed by our data in order to identify where the B-fields were accurately probed by our observations. The observations and data reduction are described in \autoref{sec:obs}. In \autoref{sec:results}, we present the polarization measurements toward IC5146. \autoref{sec:analysis} presents an initial analysis of the data to show how the dust grain alignment conditions vary with $A_V$ and regions. 
In \autoref{sec:discussion}, we discuss the evolution of dust grains and how the dust properties influence the polarization efficiency. 
The consequences of these results to the investigations of the role of B-fields in cloud evolution will be offered in the forthcoming Paper \Rom{2}.

\section{OBSERVATIONS AND DATA REDUCTION}
\label{sec:obs}
\begin{figure*}
\includegraphics[width=\textwidth]{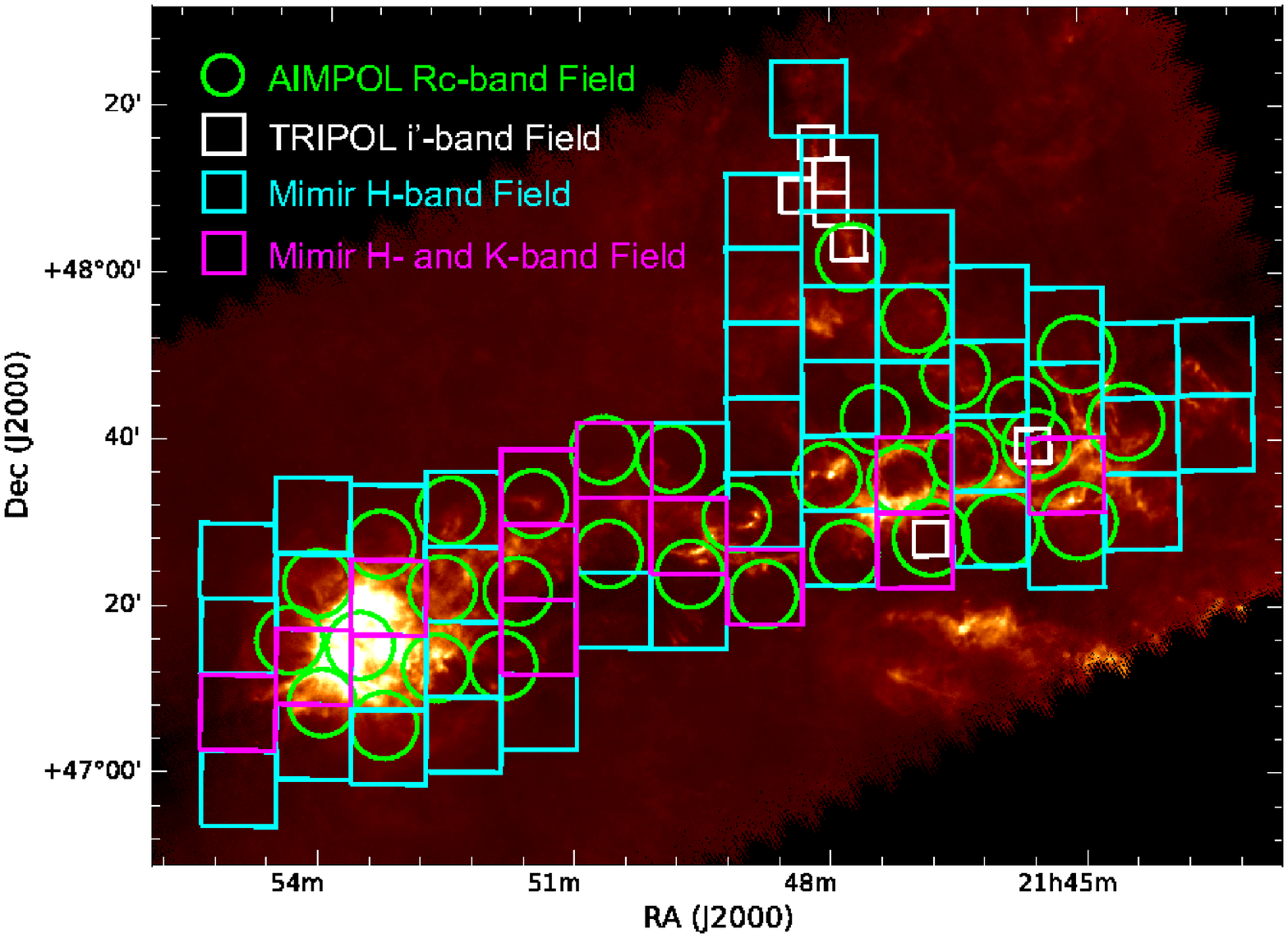}
\caption{The IC5146 fields surveyed in our polarization observations overlaid on the \textit{Herschel} 250 $\mu$m image. The fields observed with TRIPOL and AIMPOL are labeled with white boxes and green circles, respectively. The magenta boxes show the fields observed with Mimir using $H$- and $K$-band, and the cyan boxes show the fields observed with Mimir using only $H$-band. }\label{fig:obregion}
\end{figure*}

We measured the polarization of background stars toward the IC5146 filamentary cloud system with several instruments;
Aryabhatta Research Institute of Observational Sciences (ARIES) Imaging Polarimeter (AIMPOL; \citealt[]{ra04}) and Triple-Range ($g'$, $r'$, $i'$) Imager and POLarimeter (TRIPOL; \citealt[]{sa11,sa17}) provided optical $R_{c}$- and $i'$-band polarimetry data, respectively. Mimir \citep{cl07} measured the $H$- and $K$-band polarizations at near-infrared wavelengths. 
\autoref{fig:obregion} shows the target fields for each instrument overlaid on the \textit{Herschel} archive 250 $\mu$m image \citep{gr10,ar11}. 
The dust continuum map shows an east-west main filament connected to the bright cluster complex, known as the Cocoon Nebula. Numerous ``sub-filaments'' are extended from the main filament, with 
the largest sub-filament located on the north-west side of the main filament.
Almost all parts of the cloud were observed with Mimir, while the fields observed with 
AIMPOL were chosen to be mainly in the edge of the cloud in order to cover as many optically bright stars as possible. The TRIPOL observations focused on the northwestern region, with longer exposure time, where bright stars are rare both at optical and infrared wavelengths. 
The detailed information of the observations with the different instruments are described below.

\subsection{AIMPOL Polarimetry}
The $R_{c}$-band (0.67$\mu$m, bandwidth=0.14 $\mu$m) polarimetric observations  were carried out toward the IC5146 filamentary cloud system using AIMPOL mounted at the Cassegrain focus of the 104-cm Sampurnanand telescope of the ARIES, Nainital, India. AIMPOL consisted of a half-wave plate (HWP) modulator and a Wollaston prism beam-splitter illuminating a fraction (370$\times$370 pixel$^{2}$) of the Tektronics 1024$\times$1024 pixel$^{2}$ CCD camera. The pixel size of the CCD was 1.73\arcsec\ and the typical seeing was $\sim$ 2\arcsec. The observations spanned the nights of 2011 November 4--8 and 2012 November 8--15 and covered 37 fields (see \autoref{fig:obregion}) with each field having a useful diameter of $\sim$ 8\arcmin. In order to obtain Stokes Q and U, we took images at four independent HWP orientations ($0\degr$, $22.5\degr$, $45\degr$, and $67.5\degr$; \citealt[]{sc07}). 
The Wollaston prism analyzer produced offset, but simultaneous, ordinary and extraordinary overlapping images for each HWP orientation. Hence, each image resulted in one measurement of Stokes Q or U. 
The integration time was set to 10 minutes for each HWP orientation, so the total integration time for each field was 40 minutes.

The polarization measurements were calibrated for instrumental polarization and offset angle by observing standard polarized and unpolarized stars drawn from \citet{sc92}. 
The data were reduced using standard IRAF procedures \footnote{IRAF is distributed by the National Optical Astronomy Observatory, which is operated by the Association of Universities for Research in Astronomy (AURA) under a cooperative agreement with the National Science Foundation.}. The fluxes in the ordinary and extraordinary beams for each observed source were extracted using standard aperture photometry. Sources with stellar overlap, less than 10\% of the total sources, were excluded. The polarization degree and angle for each star were derived from the relative fluxes in the ordinary and extraordinary beams. We removed the Ricean bias with the asymptotic estimator 
\begin{equation}
P=\sqrt{(U^2+Q^2)-\sigma_P^2},
\end{equation}
where $P$ is the debiased degree of polarization, $Q$ and $U$ are normalized Stokes $Q$ and $U$,
and $\sigma_P$ is the uncertainty in the polarization percentage \citep{wa74}. The details of the observing facility and procedures used to estimate the polarization degree and polarization angles (P.A.) are described in \citet{es11,es12}.

\subsection{TRIPOL Polarimetry}
Polarimetric observations focused on the northwestern region of IC5146 in $i'$-bands (0.77 $\mu$m, bandwidth 0.15 $\mu$m) on 2012 July 27--28 and 2014 July 6 with TRIPOL installed on the Lulin-One-meter Telescope in Lulin Observatory, Taiwan. Seven fields, each with size of 4\arcmin$\times$4\arcmin , were observed (see \autoref{fig:obregion}). TRIPOL consisted of dichroic mirrors and three ST-9 512$\times$512 pixel$^{2}$ CCD cameras, which enabled simultaneous observations in $g'$-, $r'$-, and $i'$- bands; however, only the $i'$-band data were able to detect background stars behind this dark cloud. The pixel sizes of the CCDs were 0.5 arcsec and the average seeing was $\sim$ 1.5\arcsec. A rotatable achromatic half wave-plate and a fixed wire-grid were used to analyze the incoming light. Each field was measured at four HWP orientations ($0\degr$, $22.5\degr$, $45\degr$, and $67.5\degr$, with pairs of images yielding stellar Stokes Q and U values.). The integration time for each position angle was 22.5 minutes, so each field required a total of 1.5 hours. 

Standard reduction procedures were applied using IRAF, and the photometry of each background star was obtained using Source Extractor \citep{be96}. The debiased polarization degree and angle for each background star were derived from the fluxes measured through each of the four HWP orientation angles and calibrated against observations of polarized and unpolarized standard stars from \citet{sc92}. 

\subsection{Mimir Polarimetry}
We carried out $H$- (1.6 $\mu$m) and $K$-band (2.2 $\mu$m) polarization observations toward IC5146 on 2013 September 17--27, using the Mimir instrument \citep{cl07} mounted on the 1.8 m Perkins telescope located near Flagstaff, AZ and operated by Lowell Observatory. Fifty-seven fields covering IC5146 cloud were observed in $H$-band, and twelve fields were observed in $K$-bands toward the dense regions (see \autoref{fig:obregion}). The pixel size of the InSb detector array was 0.58 arcsec and the average seeing was $\sim$ 1.5\arcsec.
The field of view was 10\arcmin$\times$10\arcmin\ for Mimir, and we set 1\arcmin\ overlap between adjacent fields. 
Each field was observed in six sky dither positions where images were taken at 16 orientation angles of the HWP. A total of 96 (16$\times$6) images were taken for each field with 10 sec of integration time for each image for both $H$- and $K$-bands, and thus the total integration time was 16 minutes for each field. We took an additional short integration for those fields with bright stars, which took 4 minutes per field. 

In order to calibrate the non-linearities of the InSb detector array, a series of images were taken with an increasing exposure time toward an illuminated flat-field screen, and the response curve of each pixel was fitted with a polynomial model to 
obtain a linearity correction. 
Flat fields for each HWP position were taken using a lights-on/lights-off method toward a flat field screen inside the closed dome during the observation run. 
The data were calibrated using the Mimir Software Package Basic Data Processing (MSP-BDP), and the Photo POLarimetry tool (MSP-PPOL) was used to extract Stokes Q and U values for each observed source from the calibrated data. The detailed processes used in the Mimir Software Package are described in \citet{cl12a,cl12b,cl12c}. 

\section{RESULTS}
\label{sec:results}
\subsection{The Polarization Catalog}\label{sec:pcat}
We matched the polarization data to the 2MASS catalog \citep{sk06} to obtain positions accurate to 0.5 arcsec. 
Table 1 lists all the observed polarization properties as well as photometric magnitudes from 2MASS and WISE \citep{wr10}. Column 1 lists the star number, and column 2 and 3 list the RA and Dec. The measured Stokes Q, U, debiased P, and position angle P.A. in $R_c$-, $i'$-, $H$-, and $K$-bands with their uncertainties are listed in columns 4--35. The J-, $H$-, $K$-, W1-, and W2-band magnitudes and uncertainties are listed in columns 36--45. Column 46 presents the estimated visual extinction described in \autoref{sec:extinction}. Column 47 present the Usage Flag (UF; stars with UF=1 were selected for further analyses, and stars with UF=0 were not used). The usage data were selected with polarization degree divided by its uncertainty $P/\sigma_{p}$ $\geq 3$ for $R_{c}$- and $i'$-bands,
and $P/\sigma_{p}$ $\geq 2$ and $\sigma_{p}$ $<$ 5 \% for $H$- and $K$-bands; the selection criteria were relaxed for near-infrared data, since the number of 3$\sigma$ near-infrared detections were too few for adequate statistics.

In total, 2022 independent background stars have polarization detections in at least one of the four bands, 
239 stars were detected in 2 bands, 24 stars were detected in 3 bands, and only 3 stars were detected in 
all four bands. About 71$\%$ of the background stars were detected in $H$-band, 24$\%$ in $R_c$-band, 10$\%$ in $i'$-band, and 8$\%$ in $K$-band. 

\begin{figure*}
\includegraphics[width=\textwidth]{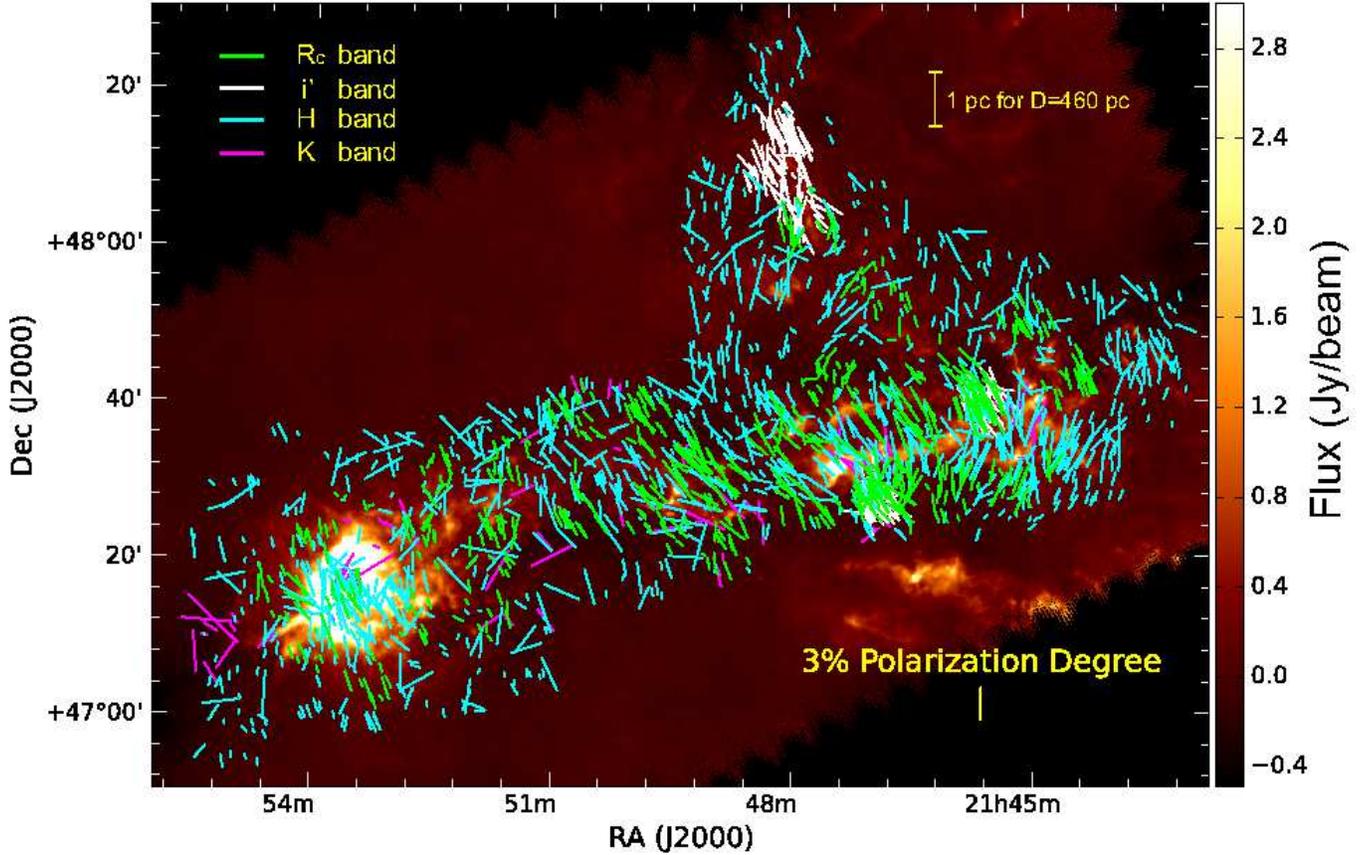}
\caption{Map of IC5146 stellar polarizations overlaid on the \textit{Herschel} 250 $\mu$m image, with FWHM beamsize of 17.6\arcsec. The detections in TRIPOL $i'$-band, AIMPOL $R_{c}$-band, Mimir $H$- and $K$-band are labeled with white, green, cyan, and magenta.}\label{fig:pmap}
\end{figure*}

\autoref{fig:pmap} shows all of the polarization measurements on the \textit{Herschel} SPIRE 250 $\mu$m image. The inferred B-field is seemingly perpendicular to the main filament on large scales but parallel to the sub-filaments. The bimodal perpendicular or parallel alignment is similar to that seen in previous polarimetry work toward filamentary clouds, and has been ascribed to B-field confinement of sub-Alfv\'{e}nic turbulence or gravitational contraction channeled by strong B-fields \citep{li13}.
We will discuss the B-field structure in detail and estimate the B-field strengths 
with the Chandrasehkar-Fermi method to determine the dynamical importance of the B-fields in Paper \Rom{2}.

\subsection{Consistency in P.A. between Multiple Wavelengths}\label{sec:cons}
\begin{figure*}
\includegraphics[width=\textwidth]{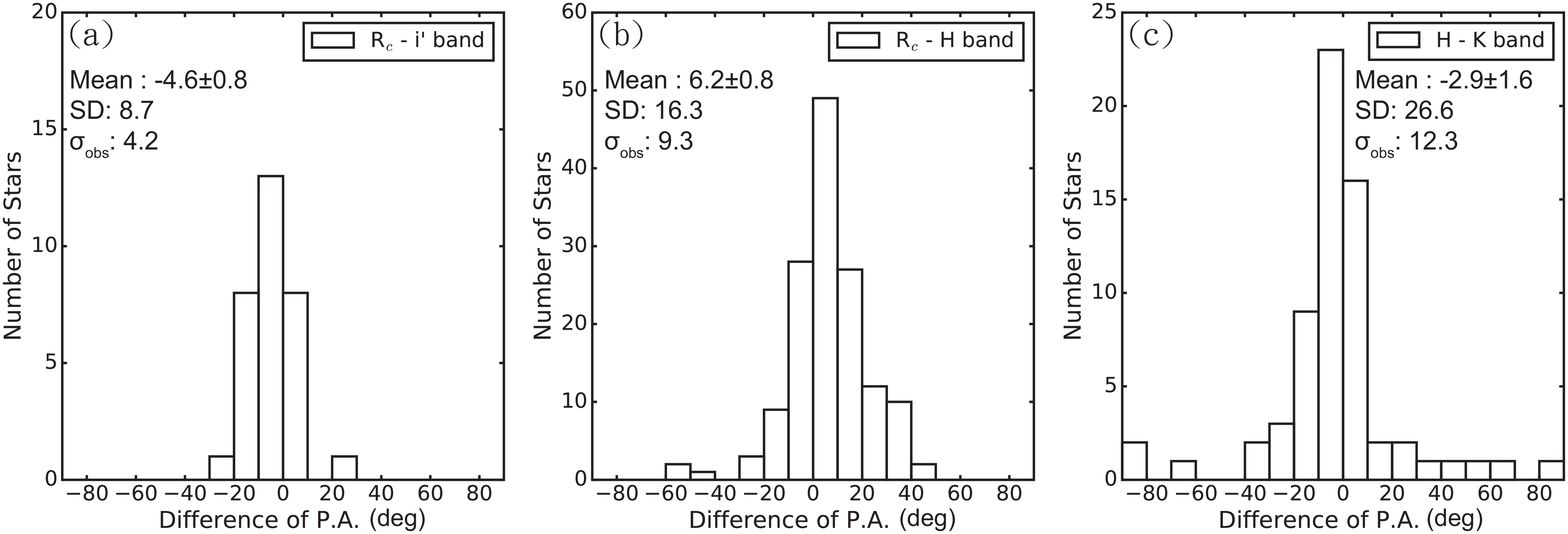}
\caption{Examination of the P.A. consistency of the data for stars with polarization detections in multiple bands. The differences of P.A.s detected in different band pairs are shown in the histograms with the mean, standard deviation, and averaged instrumental uncertainty ($\sigma_{obs}$) of the P.A.s for each band pair, listed in degrees. The comparisons between $R_{c}$- and $i'$-band number of stars, R- and $H$-band number of stars, and $H$- and $K$-band number of stars are shown in (a), (b) and (c) panels, respectively. Although small offsets and larger dispersions are shown in these band pairs compared to the instrumental uncertainties, about 90\% of the stars are distributed within 3$\sigma_{ins}$.}\label{fig:cons}
\end{figure*}

To test whether the results from AIMPOL, TRIPOL and Mimir were consistent, and thereby capable of revealing wavelength-dependent effects in the probed dust columns, 
we selected the stars with detections at multiple wavelengths, and examined their P.A.s in the different wavelength in \autoref{fig:cons}. In total, 143, 31, and 65 stars were selected in $R_c$-$i'$, $R_c$-$H$, and $H$-$K$ band pairs. The mean P.A. differences in these three band pairs were -4.6$\pm$0.8\degr, 6.2$\pm$0.8\degr, and -2.9$\pm$1.6\degr, respectively. The standard deviation of the measured P.A. differences were 8.7\degr, 16.3\degr, and 26.6\degr\ for $R_c$-$i'$, $R_c$-$H$, and $H$-$K$ band pairwise samples. The average expected uncertainties of P.A. difference, propagated from observational uncertainties ($\sigma_{P.A._{\lambda_1-\lambda_2}}=\sqrt{\sigma^2_{{P.A._{\lambda_1}}}+\sigma^2_{P.A._{\lambda_2}}}$), were 4.2\degr, 9.3\degr, and 12.3\degr\ for $R_c$-$i'$, $R_c$-$H$, and $H$-$K$ band pairwise samples. The P.A. difference standard deviations obtained from these sets are 1--2 times to the propagated instrumental uncertainties, which are acceptable values since the P.A.s may be intrinsically different at different wavelength because they may trace the polarizations to different depths. 

\subsection{Negligible Foreground Contamination}\label{sec:foreground}
\begin{figure*}
\includegraphics[width=\textwidth]{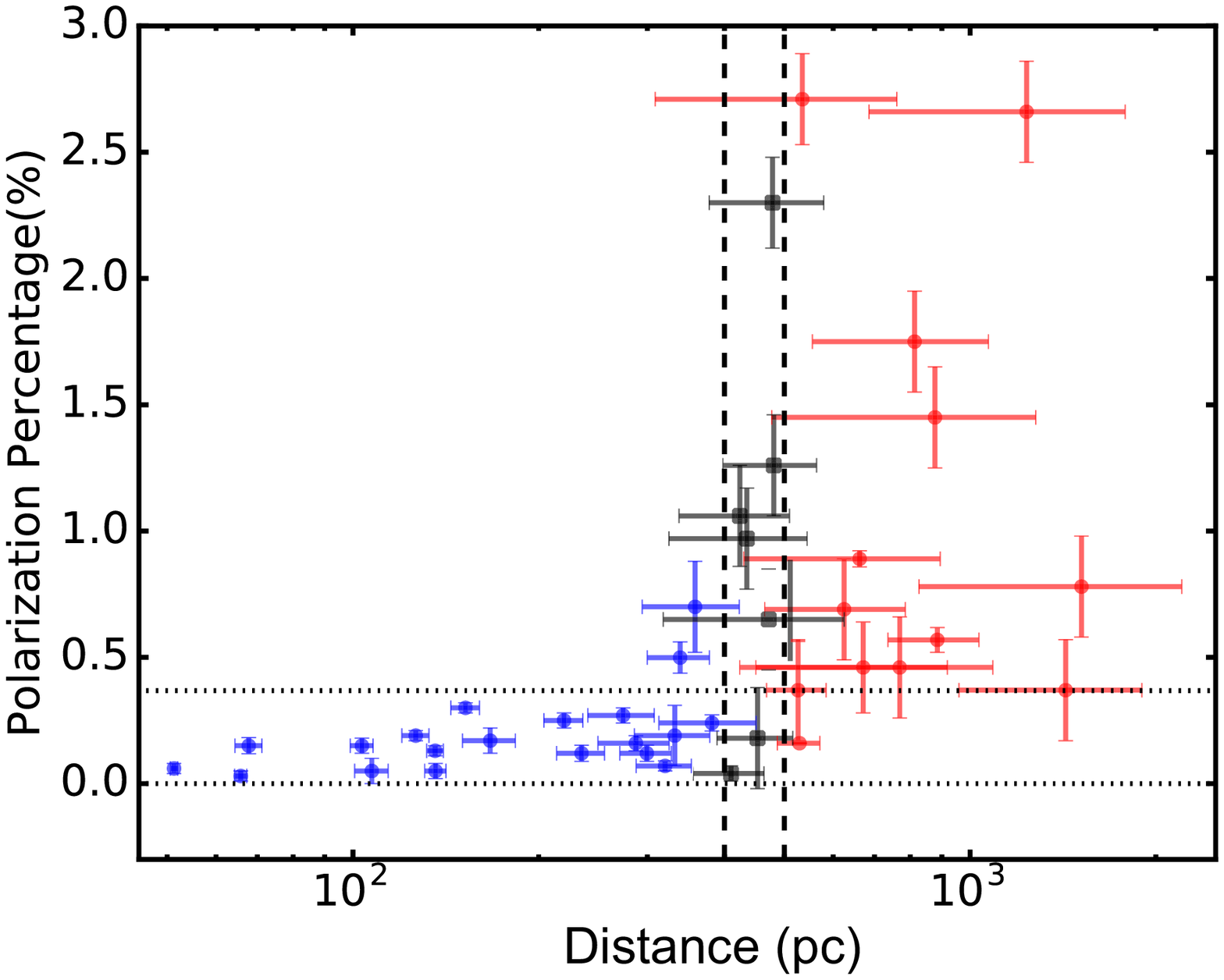} 
\caption{Polarization degree vs. distance for stars within 10 deg of IC5146 having distance measurements in the Hipparcos catalog \citep{va07} and polarization measurements in \citet{he00}. The black vertical dashed lines represent the 1-$\sigma$ upper and lower boundary of the distance of IC5146 dark cloud estimated by \citet{la99}. The blue, black, and red dots identify the stars as classfied by distance zone. The blue dot stars were used to estimate the foreground polarization to the IC5146 cloud system. The horizontal dotted lines show the estimated foreground polarization of 0--0.3\%.}\label{fig:Foreground}
\end{figure*}

We attempted to identify and exclude foreground stars whose polarizations are not tracing the B-field of the IC5146 cloud system. 
To obtain enough star samples to represent the foreground polarization near IC5146, we selected stars having known distances, and $V$-band polarization measurements from \citet{va07} and \citet{he00}, within a 10\degr\ radius sky area near IC5146. \autoref{fig:Foreground} shows the $V$-band polarization degree (P${_V}$) vs. distance for these 41 stars. The distance of $460\substack{+40 \\ -60}$\, pc to the IC5146 cloud system \citep{la99} was used to separate these stars into foreground and background groups. The polarization degree rises significantly at a distance of $\sim 400$ pc, possibly due to nearby clouds in the Gould Belt. For most of the stars with distances less than 400 pc, the polarization degrees are below 0.3$\%$. This value was chosen as the upper limit of the foreground $V$-band polarization for the sky area near IC5146. 

In the ISM, the polarization in the $V$-band is greater than in the R$_{c}$-, $H$- or $K$-bands \citep{se73}. Hence, based on the foreground star values in \autoref{fig:Foreground}, the foreground polarization in the R$_{c}$-, $H$-, and $K$-bands are expected to be less than $\sim0.3$\%. This foreground polarization upper limit is similar to the instrumental uncertainties of our data ($\sim 0.2-0.5\%$). Hence, any foreground stars were likely to have been already excluded by our selection criteria ($P/\sigma_{p}$ $>$ 3 for $R_c$- and $i'$-bands and $P/\sigma_{p}$ $>$ 2 for $H$- and $K$-bands). Thus, we concluded that contamination by foreground stars was negligible in our sample set, and further that no foreground polarization correction to the remaining data was necessary.

\section{ANALYSIS}
\label{sec:analysis}
\subsection{NICER Extinction}\label{sec:extinction}
Visual extinction provides an estimate of how many dust grains are responsible for the observed starlight polarization, and thus is required for estimating dust alignment efficiency and testing RATs theory. We used the Near-Infrared Color Excess Revisited (NICER) technique \citep{lo01}\footnote{Using the PNICER python package developed by \citet{me17}} to calculate the visual extinction over the IC5146 cloud system. 
The NICER technique utilizes multi-band colors to obtain extinction for a target field, using extinction coefficients derived from \citet{in05}. We selected a total of 186,319 stars from the 2MASS catalog and 46,100 stars from the WISE catalog covering the IC5146 cloud system. To estimate the stellar intrinsic colors, a square control field, centered at RA=330.153\degr, Dec=+47.794\degr\ with a side length of 15\arcmin, was compared to the target field. 

\begin{figure*}
\includegraphics[width=\textwidth]{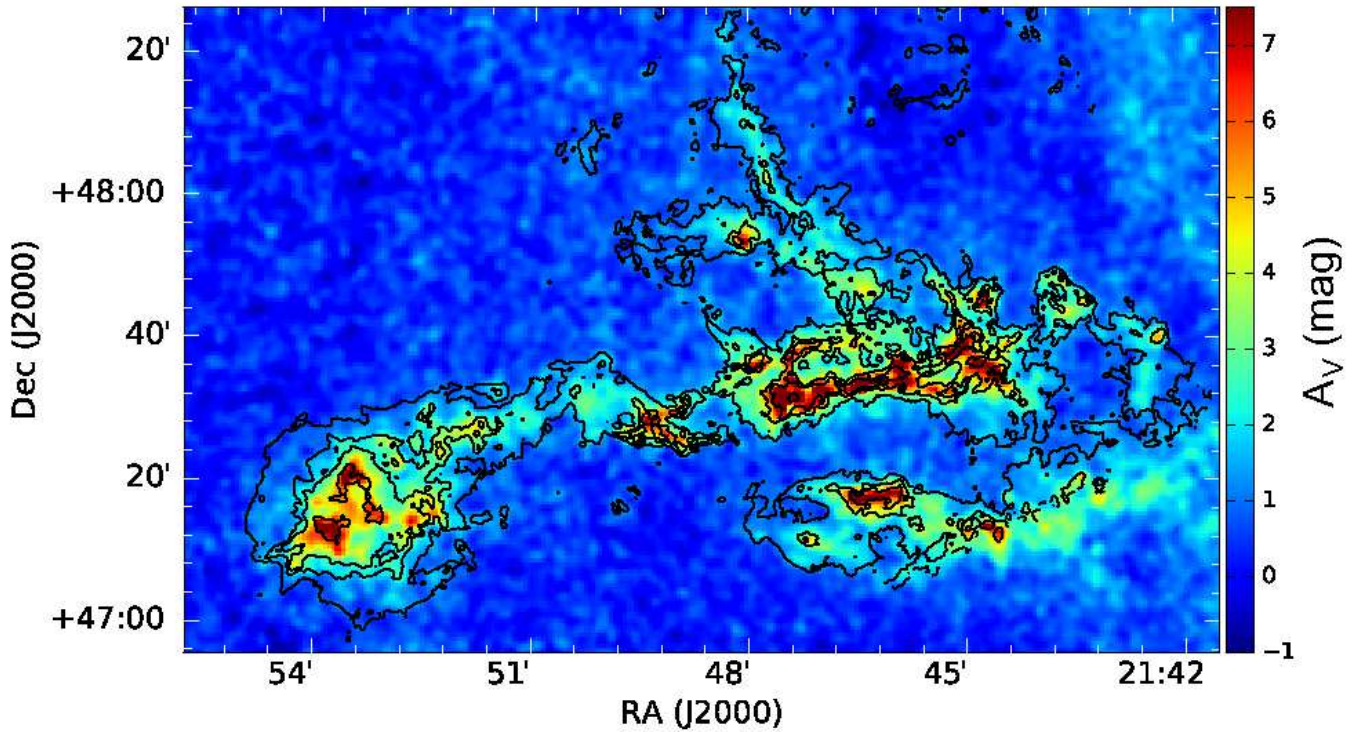} 
\caption{NICER extinction map across the IC5146 dark cloud. Visual extinction, in magnitudes of $A_V$, are displayed in color, and the black contour lines represent \textit{Herschel} 250 $\mu$m intensities with levels of 0.1, 0.5 and 1.5 mJy/beam. The FWHM of the Gaussian smoothing kernel used to create the extinction map was 90\arcsec\ and the FWHM beam size of \textit{Herschel} 250 $\mu$m image is 18.1\arcsec . The extinction morphology and 250 $\mu$m intensity distribution are in close agreement.}\label{fig:extmap}
\end{figure*}

To examine the quality of the extinctions estimated, the NICER extinction map shown in \autoref{fig:extmap} was created and compared with the \textit{Herschel} map. The extinction map was created from the variance-weighted mean of NICER extinction of each individual star within a pixel grid having a pixel size of 30\arcsec. The pixel grid was smoothed using a Gaussian weighting kernel with FWHM of 90\arcsec. The black contours in \autoref{fig:extmap} show the \textit{Herschel} 250 $\mu$m data. The morphology of the extinction map is almost identical to that of the 250 $\mu$m map.

\begin{figure*}
\includegraphics[width=\textwidth]{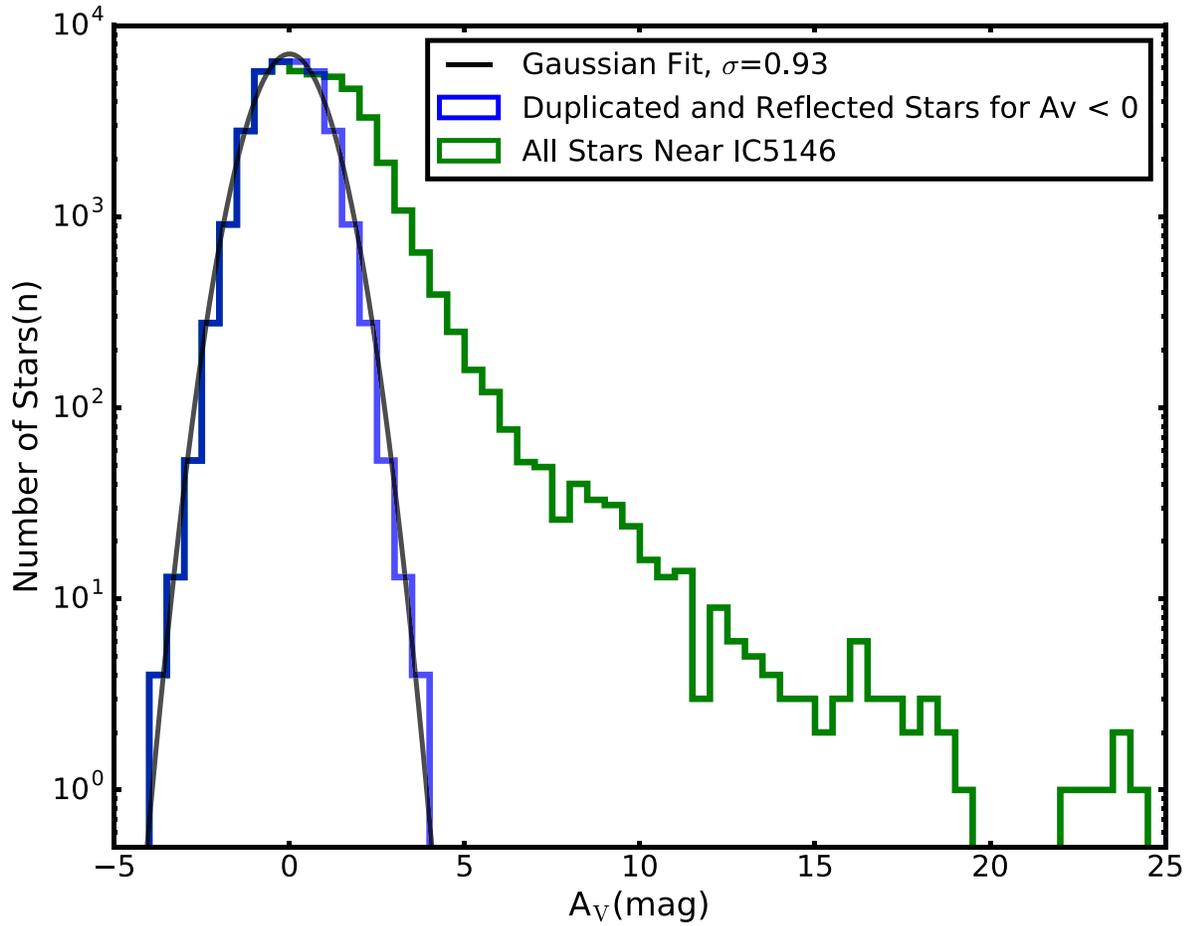}
\caption{Histogram of $A_V$ for the stars near IC5146 from the NICER analysis. In order to estimate the $A_V$ uncertainty, the negative portion of the $A_V$ distribution was duplicated and reflected about $A_V=0$ to generate the pseudo distribution shown with green. The best-fitting Gaussian, plotted with black, indicates a 1-sigma uncertainty of 0.93 mag.}\label{fig:Av}
\end{figure*}

\autoref{fig:Av} shows the histogram of $A_V$ values derived from the NICER analysis. The uncertainties in $A_V$ arise both from the uncertainties of the spectral types and from the propagated photometric uncertainties. In order to estimate the overall uncertainty of $A_V$, we selected all the stars that NICER assigned with negative $A_V$, and assumed that the negative $A_V$ values were only due to $A_V$ uncertainties. This negative $A_V$ portion of the full distribution was duplicated and reflected about $A_V$=0 to generate a new pseudo $A_V$ distribution. Fitting this with a Gaussian centered at $A_V$=0, we derived a standard deviation of 0.93 mag. Because this includes uncertainties from both spectral typing and observations, we adopted this value as our $A_V$ uncertainty.

\subsection{Polarization Efficiency}\label{sec:Peff}
To test whether the polarization measurements trace the magnetic field structure inside the IC5146 cloud system, 
we examined whether embedded dust grains align with the B-fields via estimating how the degree of polarization varied with extinction. Polarization efficiency (PE) is defined as polarization percentage divided by $A_V$. It describes how much polarization is contributed by dust grains in the line of sight. We used PE versus $A_V$ to test whether the dust grains are better aligned in the diffuse region, as predicted by the RATs model.

\begin{figure*}
\includegraphics[width=\textwidth]{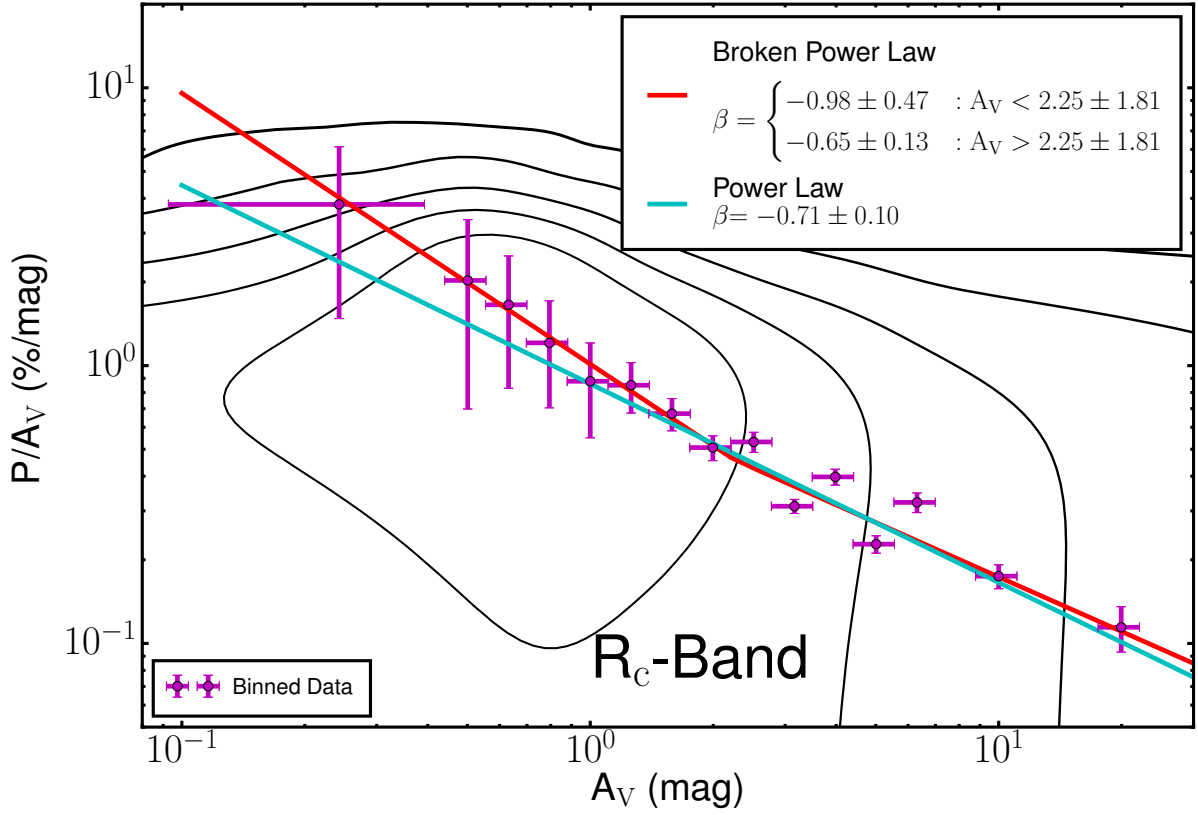}
\caption{Polarization Efficiency (PE) versus $A_V$ for $R_{c}$-band data. Black contours are drawn at 1, 10, 30, 50 and 70 percent of the peak data density. The purple points show the weighted mean PE in bins of width log ($A_V$) = 0.1. The uncertainties of PE for each bin are propagated from instrumental uncertainties, and the uncertainties of $A_V$ are propagated from 0.93 mag for each source. Those bins with $A_V <$ 0.5 mag were merged to further reduce the uncertainties. The red and cyan lines plots the best fit single power-law and broken power-law models to the binned data, respectively.}\label{fig:RPeff}
\end{figure*}

\autoref{fig:RPeff} shows PE versus $A_V$ for the $R_c$-band data. The data probability density over the PE-$A_V$ space, represented by the contours, was calculated using Kernel Density Estimation \citep{ro56,pa62}; for each source, the probability distribution function was described by a Gaussian kernel, and the width of the Gaussian kernel was determined by the uncertainties of PE and $A_V$. The probability density for the entire data set was represented by the summation of the Gaussian kernels.

To reduce the uncertainty from $A_V$ and also avoid the bias due to uneven sampling over $A_V$, the variance-weighted means of PE and $A_V$ were calculated in bins of width log($A_V$) = 0.1. The uncertainties of the weighted mean of PE and $A_V$ for each bin were propagated from the instrumental uncertainty and $A_V$ uncertainty for each sample. Those stars that NICER assigned with negative $A_V$ were not used for PE determination, since their extinctions were negligible.

\begin{figure*}
\includegraphics[width=\textwidth]{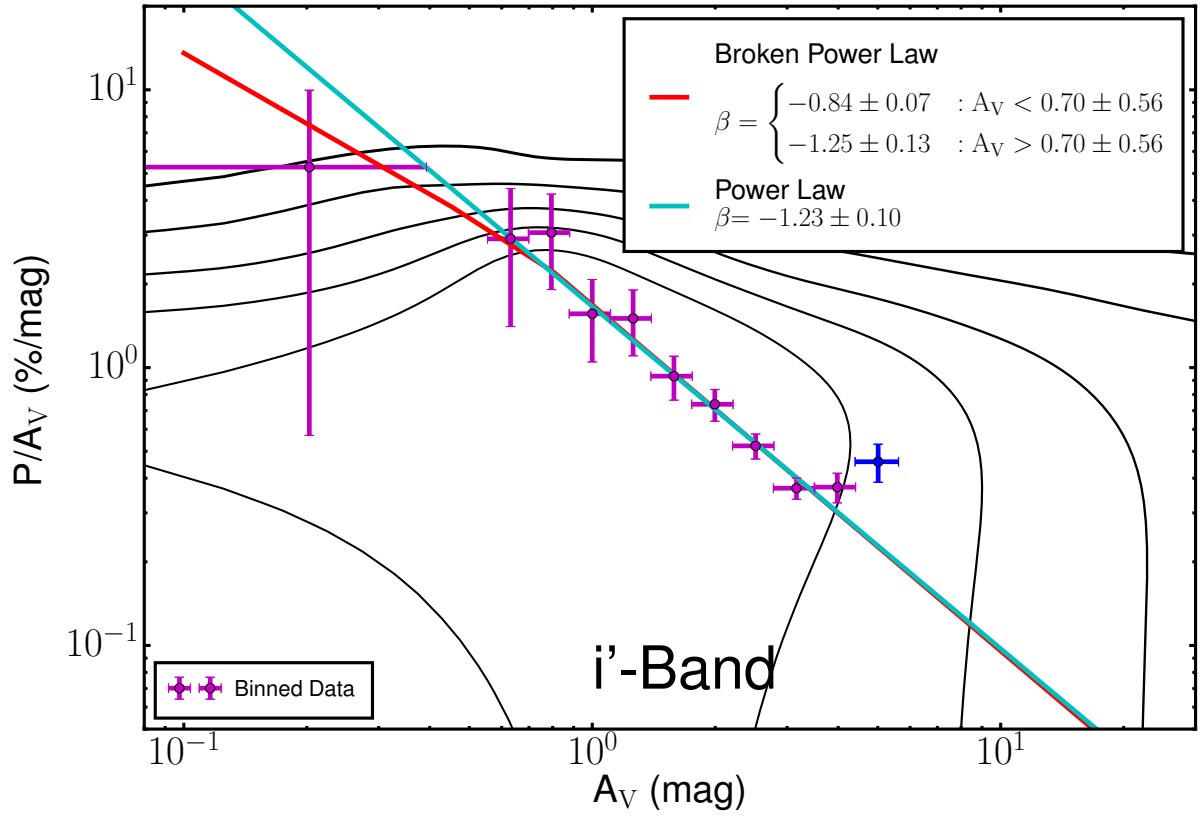}
\caption{Same as \autoref{fig:RPeff}, but for $i'$-band data. The bin with the highest $A_V$, plotted with blue, contains only two stars which seem to have uncommonly high PE, and thus it was excluded from fitting.}\label{fig:iPeff}
\end{figure*}

\begin{figure*}
\includegraphics[width=\textwidth]{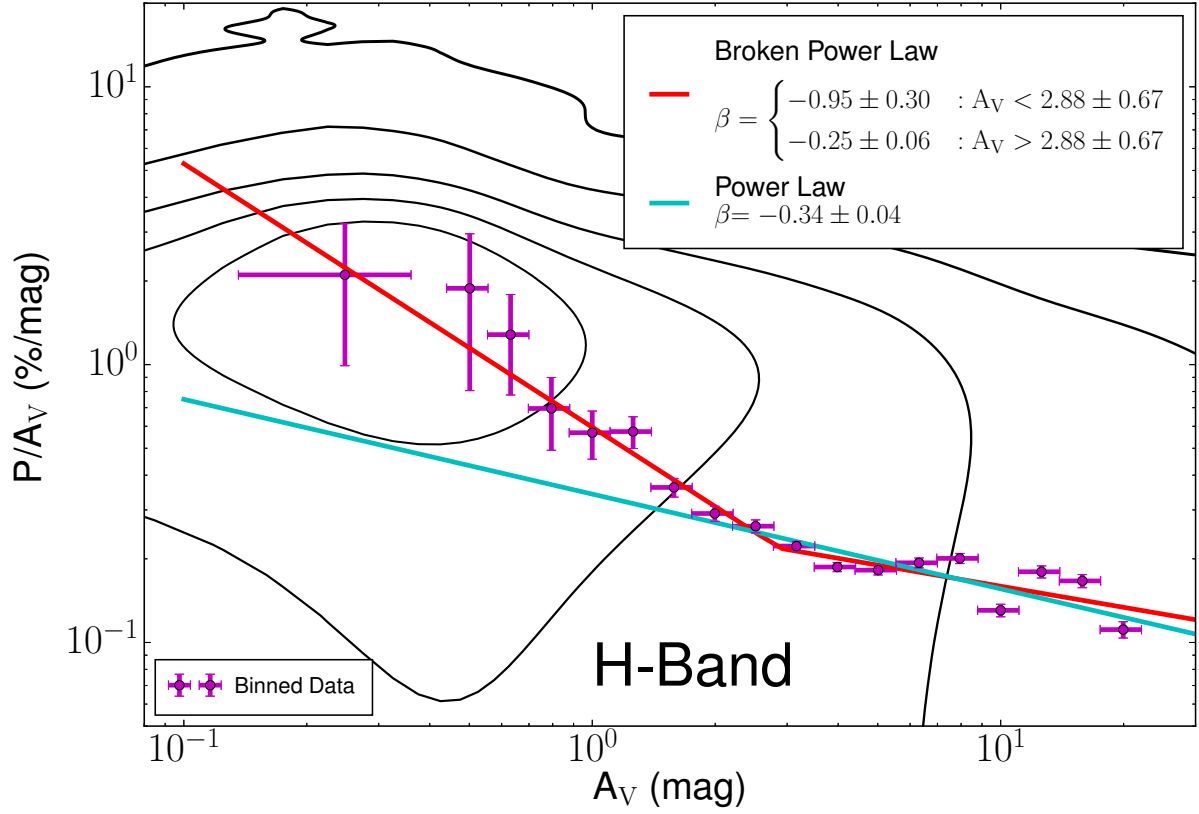}
\caption{Same as \autoref{fig:RPeff}, but for $H$-band data.}\label{fig:HPeff}
\end{figure*}

\begin{figure*}
\includegraphics[width=\textwidth]{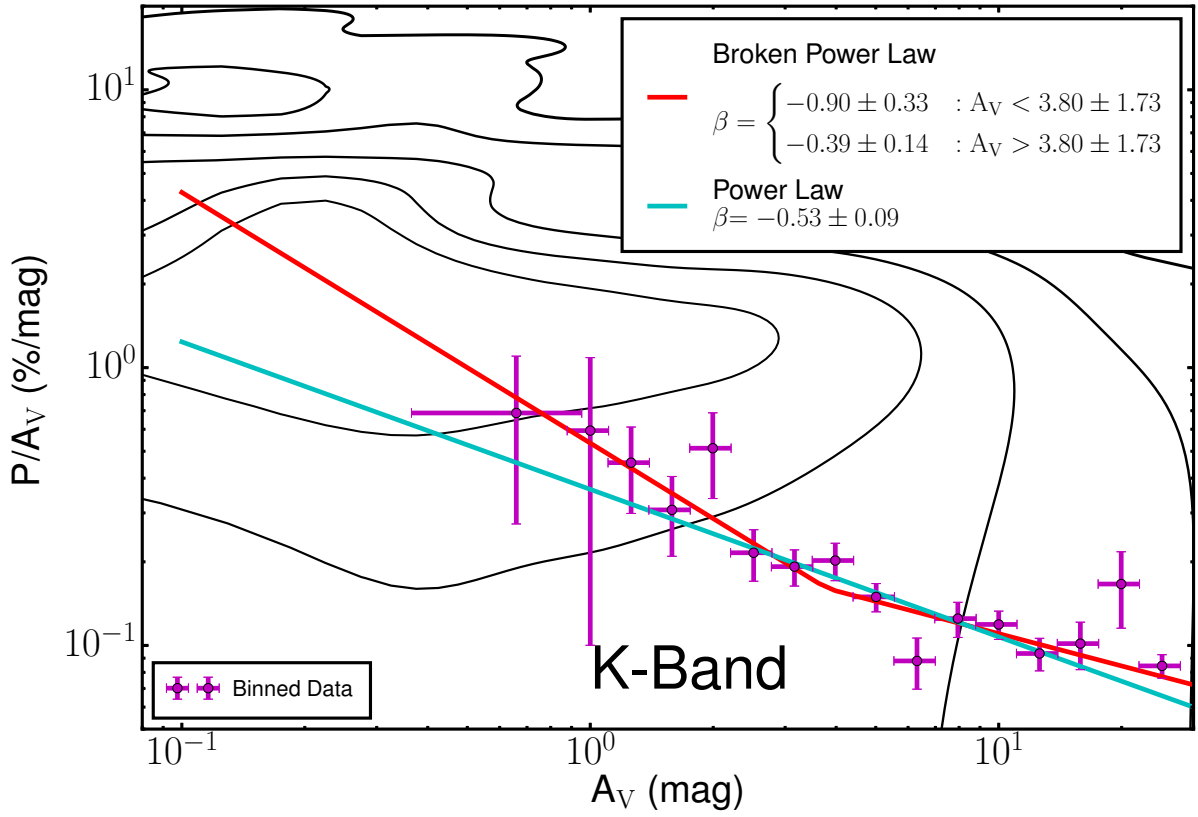}
\caption{Same as \autoref{fig:RPeff}, but for $K$-band data. Those bins with $A_V <$ 1.0 mag were merged to further reduce the uncertainties. }\label{fig:KPeff}
\end{figure*}

Similarly, PE versus $A_V$ plots for the $i'$-, $H$-, and $K$-bands are shown in \autoref{fig:iPeff}, \autoref{fig:HPeff}, and \autoref{fig:KPeff}, respectively. In $i'$-band, the last bin ($A_V$ = 5 mag) was found to have significantly higher PE and the bin only contained two stars. This bin was judged to be an outlier, and was excluded from further analysis. In the four bands, PE always decreased with $A_V$ but with different slopes. In addition, the PE at $A_V\sim 20$ mag is still half of the PE at $A_V\sim 4$ mag, suggesting that the dust grains in high $A_V$ regions are still being aligned with some degree of efficiency. 

The power-law behavior of PE-$A_V$ has been shown in previous studies \citep[e.g.,][]{wh08,ch11,ca14}, and apears to be well matched to the prediction of RATs theory. 
However, recent studies have also discovered changes in the index of the power-law for high extinction regions \citep{al14,an15}. 
To test whether, in the IC5146 dark cloud system, the power-law index changes with $A_V$,
we fit P$_{\lambda}$/$A_V$ versus $A_V$ with both a single power-law (hereafter, Model 1):
\begin{equation}
P_{\lambda}/A_V=\alpha A_V^{\beta},
\end{equation}
and with a broken power-law (hereafter, Model 2):
\begin{equation}
P_{\lambda}/A_V=
\begin{cases}
\alpha_1 A_V^{\beta_1} & : A_V \leq BP\\
 \alpha_2 A_V^{\beta_2} & : A_V > BP\\
\end{cases}
\end{equation}
where $\alpha_2$=$\alpha_1 BP^{\beta_1-\beta_2}$. The four parameters, $\alpha_1$, $\beta_1$, and $\beta_2$, and $BP$, were all taken as free parameters in the fit to the Model 2.

The goodness of fit on the binned data to the two models was examined using the F-test and the bias-corrected Akaike information criterion (AICc, \citealt[]{ak74,su78}). 
The two different model comparison methods may exhibit different preferences. For example, \citet{lu94} used Monte Carlo simulations to examine the performance of these methods, finding that the F-test tends to choose the simpler model more often than does the AICc even when the more complex model is correct. To avoid possible bias, we used both methods to compare the model fits, and the preferred Model was chosen to be the one which (1) the F-test probability is below 0.05 (95\% confidence level) and (2) has lower AICc value. The results of the fitting and model comparisons are listed in Table 2. Both the F-test and AICc show consistent results; the broken power-law is the better model for $H$-bands, and the single power-law is the better model for $R_{c}$-, $i'$-, and $K$-bands. 

The PE-$A_V$ relations were empirically found to be insensitive to observing wavelengths \citep[e.g.,][and references therein]{an15}, although our observed PE-$A_V$ relations might be expected to show some differences among the observing bands, since each data set covered different $A_V$ and spatial ranges.
The $R_{c}$- and $i'$-band data mostly probed the regions with $A_V \lesssim$ 4 mag, and thus the derived PE indices mainly characterize the low $A_V$ regions. 
The power index of $-0.71\pm0.10$ for $R_{c}$-band is very different from the index for the $i'$-band, namely $-1.23\pm0.10$. In addition, the power index of $-0.95\pm0.30$ for $H$-band for low $A_V$ values is in between the indices for $R_{c}$-band and $i'$-band. 
For high $A_V$ regions, the index of $-0.25\pm0.06$ for $H$-band is softer than the index for $K$-band ($-0.53\pm0.09$), and both are flatter than all of the indices characterizing low $A_V$ regions. 
This finding may indicate that the physical properties of grains, as well as their alignment efficiencies, change significantly beyond $A_V\sim3$--4 mag (see \autoref{sec:breakpoint})

The $R_{c}$-, $i'$-, and $H$-band data all covered the $A_V \lesssim 4$ mag regime; however, 
the power-law indices for these three bands were all different. 
One major difference between the $R_{c}$-, $i'$-, and $H$-band data was that the $H$-band observations covered almost all of the IC5146 cloud system while the $R_{c}$-band observations mostly covered the main filament, and the $i'$-band observations only probed the north-west filament (see \autoref{fig:obregion}).  

To test whether the indices characterizing low $A_V$ regions differ from region to region, the stars with $H$-band detections and $A_V < 3$ mag were assigned to one of the five zones delineated in \autoref{fig:HPeff_diffuse}. Within each zone, single power-law indices were derived and are shown in \autoref{fig:HPeff_diffuse}. Power-law indices ranging from $-0.81\pm0.13$ to $-1.80\pm0.38$ were fit to the assigned $H$-band data within the different zones. The indices were significantly softer in the eastern part of the cloud than in the western and northern part. In addition, the index derived within the northern zone was -1.18$\pm$0.14, nearly identical to the index for $i'$-band, namely $-1.23\pm0.06$, covering a substantially similar region. Hence, the PE-$A_V$ relation varies by region in the low extinction regions. The different indices derived for the different wavelengths for the whole cloud could then merely arise from the combination of a variety of PE-$A_V$ relations. The possible origin of the diverse PE relation is discussed further in \autoref{sec:diPE}.

\subsection{Wavelength Dependence of Polarization using the Serkowski relation}\label{sec:Pspec}
The polarization of starlight is known to be wavelength-dependent, resulting in a polarization spectrum. Since this wavelength-dependence originates from dust properties and alignment conditions, the polarization spectrum can be used to investigate the evolution of dust grains.
The polarization spectrum is well-fit by the empirical ``Serkowski relation,'' 
\begin{equation}
P(\lambda)=P_{max}\cdot exp\{-K\cdot ln^2\left(\lambda_{max}/\lambda\right)\}
\end{equation}
where $P_{max}$ is the peak polarization degree at wavelength $\lambda_{max}$\ \citep{se73}. 
In later studies, the parameter K was shown to follow the relation $K= -0.1 +$ c$\lambda_{max}$ \citep{wi82}, where c was found to depend on dust properties, such as geometric shape \citep[e.g.,][]{vo13,vo14}.

To fit the Serkowski relation for its three parameters, detections in at least four bands are required to obtain uncertainties in the parameters. 
However, it is difficult to do so in the presence of high extinction. Thus, most of the previous studies have been limited to low extinction regions. 
Even among our data, only three stars were detected in all four wavebands. 
The Serkowski relation fit results for these stars are shown in \autoref{fig:Serkowski_Fit}. The range of fitted $\lambda_{max}$ is $\sim$ 0.72--0.83 $\mu$m, significantly greater than the typical value $\sim 0.55$ $\mu$m characterizing the general diffuse interstellar medium \citep{se75}.

\subsection{``Polarization color" as a constraint on $\lambda_{max}$}\label{sec:Pcolor}
It is difficult to obtain the necessary multiple wavelength polarization detections needed to fit Serkowski relations in the presence of strong extinction. 
In order to have more samples to constrain $\lambda_{max}$, we examined the ``polarization color'' (ratio of polarization degrees at two bands) as a partial descriptor of the polarization spectrum. 
Polarization color can be easily obtained from a limited set of wavebands without presumption of the spectrum shape. The Serkowski relation has been suggested to be only valid for a limited wavelength range, mostly UV through optical. \citet{ma89} found the polarization spectrum between 1.6--5 $\mu$m wavelength behaved more like a power-law than like the Serkowski relation, and \citet{cl92} found the polarization spectrum in the ultraviolet showed an excess with respect to the Serkowski relation. 

In the wavelengths where the Serkowski relation is valid, polarization color can constrain $\lambda_{max}$. 
Following the Serkowski relation, the ratio of polarization degree of a star at two wavelengths can be written as:
\begin{equation}\label{eq:PPR}
\frac{P(\lambda_1)}{P(\lambda_2)}=\frac{exp\{-K\cdot (ln^2\left(\lambda_{max}/\lambda_1\right))\}}{exp\{-K\cdot (ln^2\left(\lambda_{max}/\lambda_2\right))\}}
\end{equation}
To reduce the free parameters, we also assumed the relation K= $-0.1 +$ c$\lambda_{max}$, where c is expected to be a constant over $A_V$. 
Via \autoref{eq:PPR} and the assumptions regarding K, polarization color (ratio) is determined by only one free parameter, $\lambda_{max}$. Thus, only one polarization color is required to constrain $\lambda_{max}$, although for wavebands right near $\lambda_{max}$, the uncertainties will be large. \citet{cl16} used this approach with Mimir $H$- and $K$-band data to identify grain growth in the moderate to high extinction regions of L1544, finding $\lambda_{max}$ in the 1.0-1.2 $\mu$m range.

\begin{figure*}
\includegraphics[width=\textwidth]{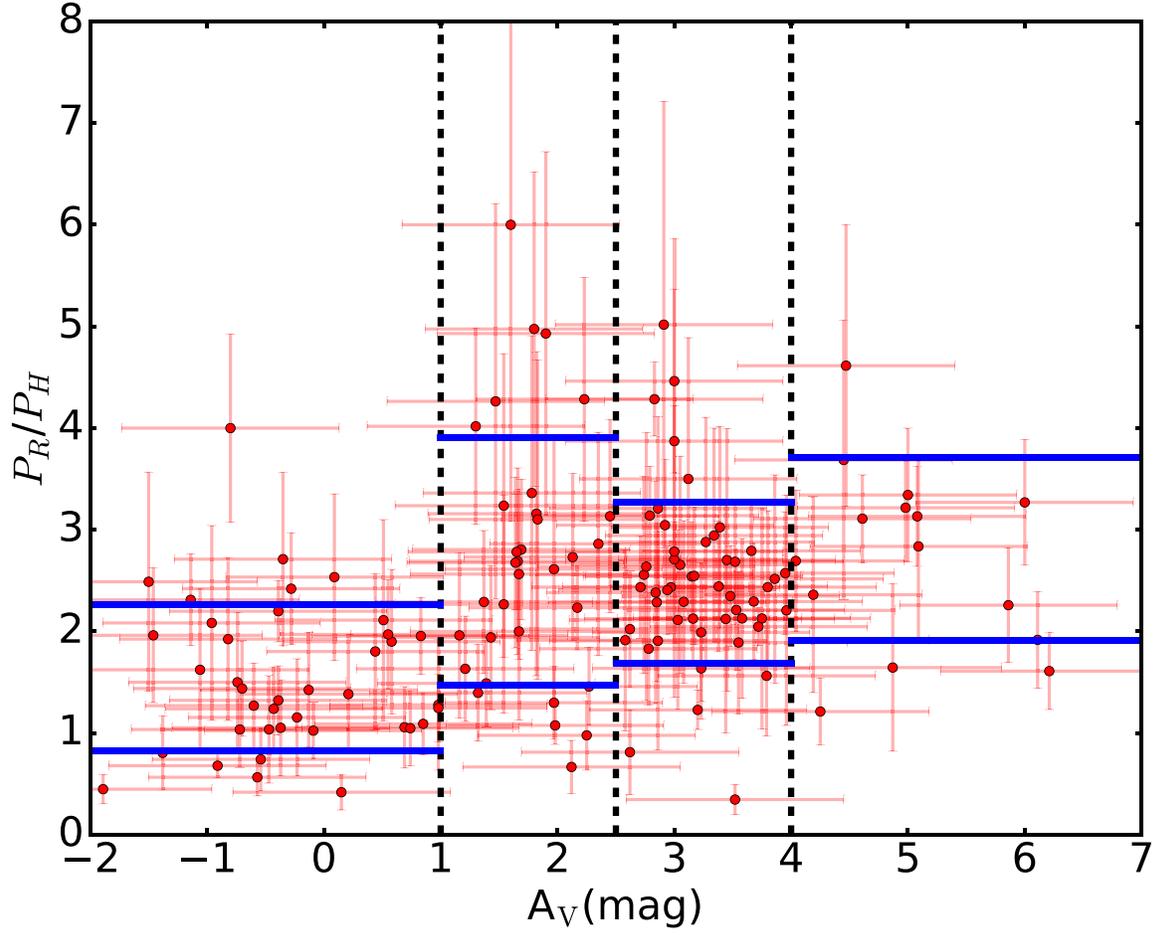}
\caption{P$_{R_c}$/P$_H$ vs. $A_V$ from the stars with $R_c$- and $H$-band polarization detections. These stars were separated into 4 groups, based on their $A_V$ values as defined by the black dashed lines. The blue horizontal lines show the $\pm1\sigma$ excursions above and below the average of P$_{R_c}$/P$_H$ for each group. The distribution of P$_{R_c}$/P$_H$ is narrowest when $A_V$ = 2.5--4 mag. The stars with negative $A_V$ values have negligible extinction: these values result from uncertainties in the NICER extinction estimation.}\label{fig:RH_ratio}
\end{figure*}

The polarization color $P_{R_c}/P_{H}$ versus $A_V$ values shown in \autoref{fig:RH_ratio} reveal how the polarization spectrum varies with $A_V$. The observed $P_{R_c}/P_{H}$ distribution seems diverse and could be hardly described with a simple function. 
To try to find some order, we separated the stars into four $A_V$ groups by eye, based on the similarities in the distribution. They span $A_V$ ranges of $< 1$, 1.0--2.5, 2.5--4.0, and $>$ 4.0 mag, labelled as regions ``A'', ``B'', ``C'', and ``D'', respectively. The blue horizontal lines show the unweighted average$\pm$standard deviation (STD) of $P_{R_c}/P_{H}$ for each group. The intrinsic dispersions of $P_{R_c}/P_{H}$ were derived using $\sigma_{intrinsic}=\sqrt{STD^2-\sigma^2_{obs}}$, where $\sigma_{obs}$ was the average observational uncertainty. The statistical properties of $P_{R_c}/P_{H}$ for each group are presented in Table 3.

Group A had significantly lower mean $P_{R_c}/P_{H}$ than the other regions. Since the $A_V$ of this region was comparable to expected foreground extinction $<$0.3 mag \citep{la94}, the depolarization due to any foreground medium might significantly affect the polarization spectrum. Thus, the polarization color in this group might not accurately trace the dust properties. 
The average $P_{R_c}/P_{H}$ values for group B, C, and D were similar to each other, although significant intrinsic dispersion was found for group B.

\begin{figure*}
\includegraphics[width=\textwidth]{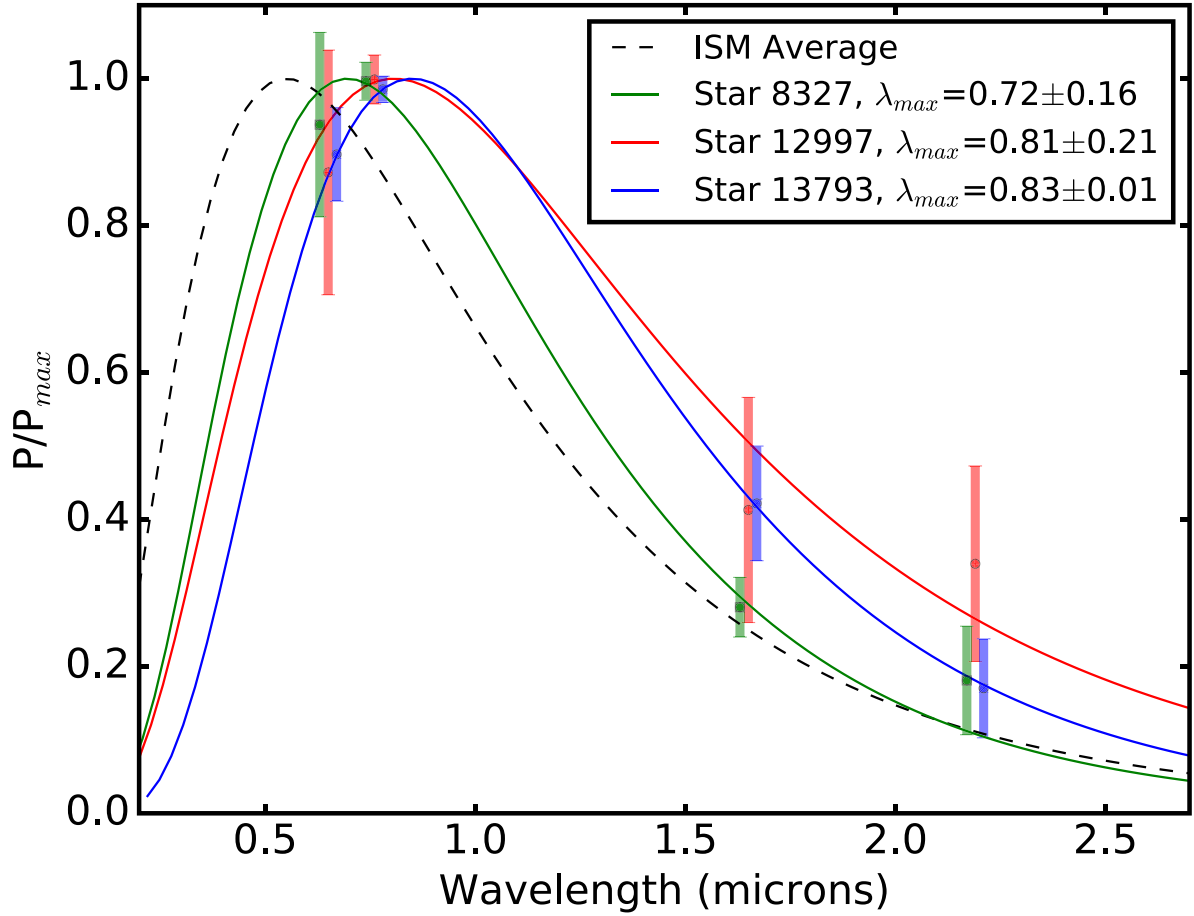}
\caption{Polarization spectra for the three stars with polarization detections in 4 bands. Colored curves overplot the best fit Serkowski relation. Wavelengths of the data points and error bars for the Star 8327 and 13793 were slightly shifted to avoid overlapping. The black dashed spectrum represents the typical ISM average with $\lambda_{max}$ of 0.55 $\mu$m. All three stars have larger $\lambda_{max}$ values than the ISM average. }\label{fig:Serkowski_Fit}
\end{figure*}

\begin{figure*}
\includegraphics[width=\textwidth]{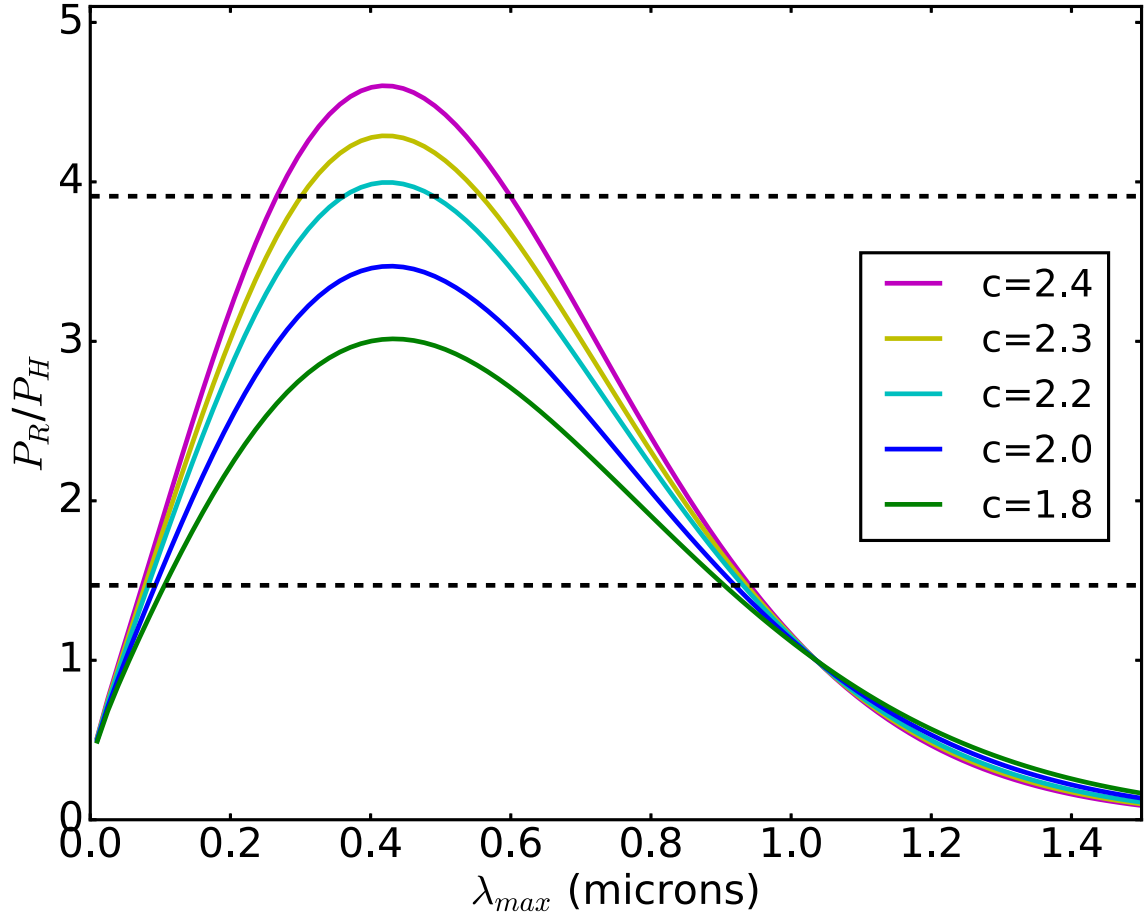}
\caption{Relations between P$_{R_c}$/P$_H$ and $\lambda_{max}$ predicted by the Serkowski relation assuming $K=c\dot \lambda_{max}$. The horizontal dashed lines represent the range of observed P$_{R_c}$/P$_H$, seen as the range spanned by the blue horizontal lines shown in \autoref{fig:RH_ratio}. Only the relation characterized by c $\geqslant$ 2.2 can cover the range of observed P$_{R_c}$/P$_H$. The yellow curve shows the relation for c$=2.3$, which was used to convert P$_{R_c}$/P$_H$ to $\lambda_{max}$. }\label{fig:RH_Serkow}
\end{figure*}

The value of $P_{R_c}/P_{H}$ directly determines the value of $\lambda_{max}$ if the parameter c in \autoref{eq:PPR} is known. The variance-weighted mean c of 2.3 was derived from the four-band fitting to the three stars shown in \autoref{fig:Serkowski_Fit}. This c value was higher than the value of 1.66 found in \citet{wi82} and the value of 1.86 found in \citet{wh92}. \autoref{fig:RH_Serkow} shows how $P_{R_c}/P_{H}$ maps to $\lambda_{max}$ for different value of c. The black horizontal dashed lines delimit the span of the distribution of $P_{R_c}/P_{H}$ values found in \autoref{fig:RH_ratio}. The curves in \autoref{fig:RH_Serkow} characterized by c $\leqslant$ 2.1 cannot cover the full range of the distribution of observed $P_{R_c}/P_{H}$ values, suggesting that the parameter c value in IC5146 cloud system is likely higher than it is in other clouds. Hence, we assumed c $=$ 2.3 to estimate the $\lambda_{max}$ from observed $P_{R_c}/P_{H}$.

\begin{figure*}
\includegraphics[width=\textwidth]{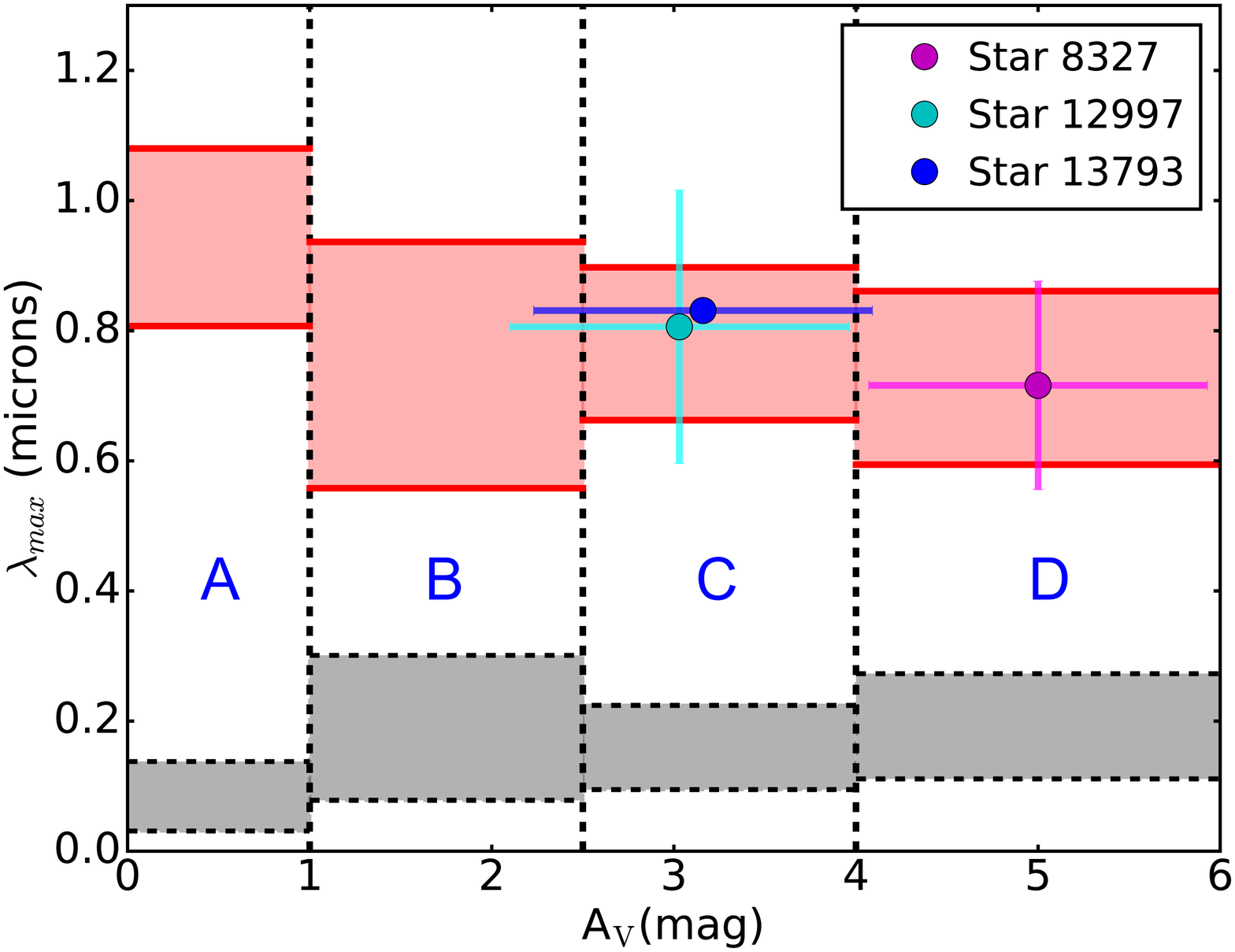}
\caption{Ranges of $\lambda_{max}$ with $A_V$ group. The colored regions represent the ranges of the distributions of $\lambda_{max}$ values in each $A_V$ group, estimated from the observed P$_{R_c}$/P$_H$ values, the Serkowski relation, and the c $=$ 2.3 curve from \autoref{fig:RH_Serkow}}. The red and grey sets of zones identify the two degenerate sets of solutions to the Serkowski relation. The three stars whose four-band polarimetry data were fitted to the Serkowski relation (see \autoref{fig:Serkowski_Fit}) are plotted as filled, colored circles with error bars and are consistent with the set of red zone solutions corresponding to greater values of $\lambda_{max}$.\label{fig:RH_lambda}
\end{figure*}

The range of $\lambda_{max}$ versus $A_V$, for c $=$ 2.3, is shown in \autoref{fig:RH_lambda}. The mean $\lambda_{max}$ for the mean $P_{R_c}/P_{H}$ in each of the four $A_V$ groups is listed in Table 3. 
For each $P_{R_c}/P_{H}$, two possible $\lambda_{max}$ solutions from the Serkowski relation are possible, as can been seen in \autoref{fig:RH_Serkow}. These two solution regions were colored in gray regions for the lesser $\lambda_{max}$ values and in red for the greater $\lambda_{max}$ values. If we assume that $\lambda_{max}$ across the cloud system does not dramatically change within a few magnitudes of extinction, then only one set of solutions is likely to be true. Since the values of $\lambda_{max}$ derived from the four-band fitting to the three stars shown in \autoref{fig:Serkowski_Fit} were all in the range of 0.72--0.83 $\mu$m, $\lambda_{max} >$ 0.4 $\mu$m selects the better solutions.

Hence, the average $P_{R_c}/P_{H}$ values correspond to $\lambda_{max}$ of 0.75, 0.78, and 0.73 $\mu$m for groups B, C, and D, respectively. These $\lambda_{max}$ are all greater than the average value for the ISM of 0.55 $\mu$m \citep{se75}, suggesting that the dust grains across most of the IC5146 cloud system have grown significantly with respect to dust grains in the diffuse ISM. In addition, the large dispersion of $\lambda_{max}$ in group B implies that the dust grain size distributions are more diverse in this $A_V$ range than in the other extinction regions.

\begin{figure*}
\includegraphics[width=\textwidth]{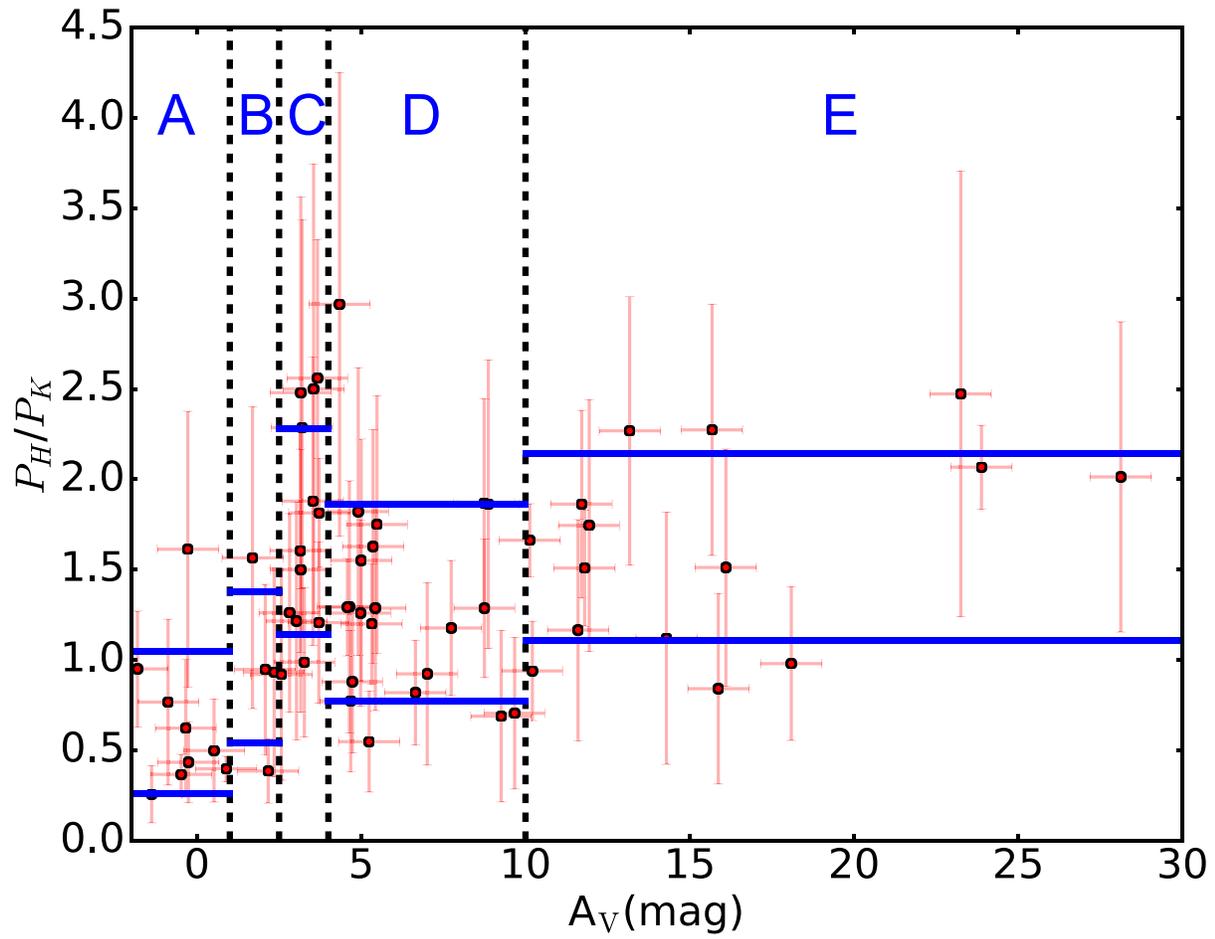}
\caption{Same as \autoref{fig:RH_ratio}, but for P$_H$/P$_K$ vs. $A_V$. The dispersions of P$_H$/P$_K$ are almost the same over the different $A_V$ groups while the P$_H$/P$_K$ means appear to change with $A_V$.}\label{fig:HK_ratio}
\end{figure*}

Similarly, we plotted $P_{H}/P_{K}$ versus $A_V$ in \autoref{fig:HK_ratio} to trace the variation of the polarization spectrum in the infrared. The data were separated into five groups, in a fashion similar to \autoref{fig:RH_ratio} ($A_V < 1$, 1.0--2.5, 2.5--4.0, 4.0--10.0, and $>$ 10.0 mag, labelled as A, B, C, D, and E, respectively). The statistical properties of $P_{H}/P_{K}$ for these groups are listed in Table 4. The average $P_{H}/P_{K}$ changes significantly with $A_V$, with the greatest value being found in group B.

The values of $\lambda_{max}$ converted from $P_{H}/P_{K}$ were listed in Table 4. The range of $\lambda_{max}$ derived from the group averaged $P_{H}/P_{K}$ was 1.46--1.93 $\mu$m, significantly higher than the  0.73--0.78 $\mu$m range derived from $P_{R_c}/P_{H}$. The inconsistency implies that the Serkowski relation may not well describe the polarization spectrum at near-infrared wavelengths. \citet{ma89} suggested that the near-infrared polarization spectrum can be better described by a power-law 
\begin{equation}\label{eq:Plambda_infrared}
P(\lambda)\propto{\lambda}^{-\beta}
\end{equation}
where $\beta\simeq$ 1.6--2.0. In the infrared, polarization color can still constrain the power-law index by
\begin{equation}\label{eq:PPR_infrared}
\frac{P_H}{P_K}=0.74^{-\beta}.
\end{equation}

The $\beta$ derived from our data are listed in Table 4. These indices probably vary with $A_V$ and have a peak at $A_V \approx$ 2.5--4 mag, as shown by \autoref{fig:HK_ratio}. \citet{ki95} showed that the index could depend on the amount of micron-size dust grains. Our results hint that the population of micron-size dust grains might evolve with $A_V$, and a significant change occurs at $A_V$ of 2.5--4 mag.

\section{DISCUSSION}
\label{sec:discussion}
\subsection{Evolution of Dust Grains}\label{sec:grain}
The derived $\lambda_{max}$ values obtained in previous studies \citep[e.g.,][]{wh78,wi82} are expected to be related to the small-size cutoff of the grain size distribution. \citet{wh78} found that $\lambda_{max}$ is related to the ratio of total to selective extinction ($R_V=A_V/(E_{B-V})$) by an empirical relation $R_V=(5.6\pm0.3)\lambda_{max}$, tracing the reddening changes with grain size distribution changes. However, with more observations, \citet{an07} found no correlation between $\lambda_{max}$ and $R_V$ for individual clouds; the empirical relation could only be recovered by combining the data for all of their observed clouds. They further argued that $\lambda_{max}$ may depend on both grain size distributions and dust alignment conditions. 

\citet{wh08} showed, in their RATs simulations with constant grain size distribution, that $\lambda_{max}$ increases with $A_V$ because small dust grains become less aligned due to the decaying and reddening of the external radiation field by extinction. 
Their model predicts a smooth increase of $\lambda_{max}$ from 0.45 to 0.75 $\mu$m as $A_V$ increases from 0 to 6 mag. However, our derived $\lambda_{max}$ values, 0.73--0.78 $\mu$m, do not significantly change with $A_V$. In addition, our data show significant decrease of the intrinsic dispersion of $\lambda_{max}$ from 0.11$\pm$0.08 ($A_V=$1.0--2.5 mag) to $<$0.07 ($A_V=$2.5--4.0 mag) possibly due to changes of the grain size distribution.

The change in the dispersion of $\lambda_{max}$ implies that the grain size distribution in the dense regions could be more uniform than in the diffuse regions. We speculate that the difference is due to the evolution of dust grain size. In grain-grain collision models \citep{jo96,hi09,or09}, the grain size distributions are expected to be modified by the competing effects of fragmentation and coagulation. Steady-state grain size distributions are eventually reached when these two effects achieve equilibrium. As a result, the apparent uniform grain size distribution in the dense regions may hint at the existence of stabilized grain-grain collision processes. Together with the increase of $\lambda_{max}$, this non-evolution of grains in the dense regions favors the notion that grain growth has already taken place before $A_V$ reaches 2.5-4.0 mag.

\begin{figure*}
\includegraphics[width=\textwidth]{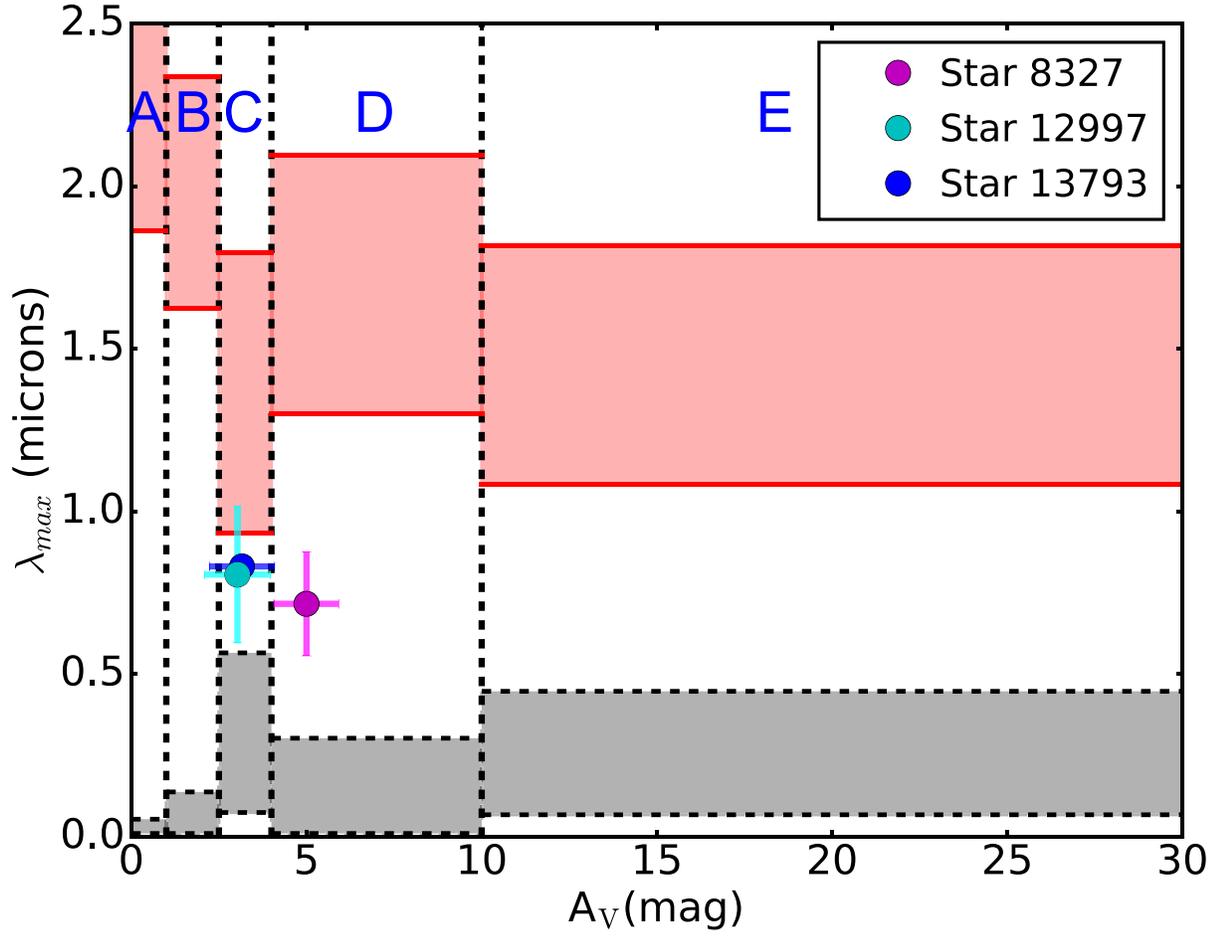}
\caption{(a) Same as \autoref{fig:RH_lambda}, but the $\lambda_{max}$ range were estimated from P$_H$/P$_K$ using the Serkowski relation with c $=$ 2.3. The $\lambda_{max}$ values derived from P$_H$/P$_K$ are much greater than the $\lambda_{max}$ values derived from P$_{R_c}$/P$_H$ shown in \autoref{fig:RH_lambda}. In addition, the distribution of $\lambda_{max}$ are inconsistent with the best fit Serkowski relations to the three stars with four-band polarimetry data shown in \autoref{fig:Serkowski_Fit}, suggesting that the Serkowski relation may not be valid, at least in $K$-band.}\label{fig:HK_lambda}
\end{figure*}

In \autoref{fig:HK_lambda}, we show that the $\lambda_{max}$ derived from a Serkowski relation based on analysis of $P_{H}/P_K$ yielded an average $\lambda_{max}$ of 1.46--1.93 $\mu$m. This was inconsistent with $\lambda_{max}$ derived from $P_{R_c}/P_H$, which yielded an average $\lambda_{max}$ of 0.73--0.78 $\mu$m. To reach a $\lambda_{max}$ of 0.73--0.78 $\mu$m, $P_{H}/P_K$ would need to be $>$ 2.3, which is much higher than the observed range of mean values from 0.96$\pm$0.28 to 1.71$\pm$0.21. This indicates that at least the measured $K$-band infrared polarization exceeds the predictions of the Serkowski relation. This excess infrared polarization was also found in early studies \citep[e.g.,][]{ma90,na90,jo90}, and is better fit by a power-law (\autoref{eq:PPR_infrared}). We found that a power-law with $\beta$ ranging from $-0.14\pm0.69$ to $1.78\pm0.41$ can explain our data for different $A_V$. The $\beta$ index appears to vary with $A_V$, and has its highest value for $A_V=$ 2.5--4.0 mag. Our $\beta$ are similar to, or smaller than, the value of $1.6\pm0.2$ measured toward ScO-Oph and CyG OB2 \citep{ma92}, and $1.76\pm0.25$ measured toward the Galactic center \citep{ha13}. 

Two models have tried to explain the physical origin of the excess infrared polarization over the Serkowski relation. 
\citet{ki95} showed that a model with a dust grain mixture with sizes of 0.3 $\mu$m and 0.6--1 $\mu$m can reproduce the observational Serkowski relation while also exhibiting a power-law excess infrared polarization. 
\citet{li97} showed that an organic refractory mantled dust grain model with a Gaussian grain size distribution could reproduce the observed excess infrared polarization without requiring a high abundance of micron-size dust grains. 
Our results show that the observed power-law index might vary with $A_V$, providing new constraints for future models.

\subsection{How Deep Into a Cloud Can Polarization Be Used to Reveal B-fields?}\label{sec:deep}
Polarization efficiency has been used to indicate how deep into a cloud, in $A_V$, the dust grains remain aligned. \citet{go95} found a PE-$A_V$ power-law index of $-1$ in L1755 in the JHK bands. This constancy of P with $A_V$ was used to argue that near-infrared polarization does not trace the B-field within dense clouds. In contrast, \citet{wh08} found an index of $-0.52$ in Taurus at $K$-band, showing that polarization can trace the B-field, at least up to $A_V \sim 10$ mag. 

The power-law PE versus $A_V$ indices characterizing these two cases can be found in our data, but for different $A_V$ ranges. 
The power index of $-0.95\pm0.30$ measured for $H$-band in low $A_V$ regions is similar to the results found in L1755 at $R$-band \citep{go95} for $A_V <$ 8--10 mag . 
Our index of $-0.53\pm0.09$ for $K$-band is close to the value found in Taurus at $K$-band ($-0.52\pm0.07$) by \citet{wh08} and the value found in numerous starless cores at $K$-band ($\sim -0.5$) for $A_V <$ 20 mag by \citet{jo15}. In addition, the intermediate index $-0.71\pm0.10$ found in the $R_c$-band is similar to the results found in Pipe-109 using $R$-band ($-0.76\pm0.14$, \citealt{al14}) and in L204 using $H$-band ($-0.74\pm0.07$, \citealt{ca14}). We found our steepest index, of $-1.23\pm0.10$, in the $i'$-band and our softest index, of $-0.25\pm0.06$, in the $H$-band in high $A_V$ regions.

The softer indices of $-0.25$ to $-0.53$ are only found in high $A_V$ regions in $H$- and $K$-bands, while a variety of steeper indices are found in low $A_V$ regions. The $\sim -0.5$ indices match predictions of RATs models with constant grain size distributions \citep{wh08}. The same models predicts a steepening of PE as $A_V$ approaches 10 mag, due to the strong extinction of external radiation. Here, however, the indices we measure for high $A_V$ regions show that the dust with $A_V$ up to $\sim$ 20 mag still contributes to the measured polarization: the polarization efficiency at $A_V=20$ mag is only a factor of $\sim$ 2 lower than that at $A_V=$4 mag. Thus, the notion that only the dust on the surfaces of clouds is aligned, as concluded by \citet{go95}, is not supported here.

\subsection{The Diverse PE in Low $A_V$ Regions}\label{sec:diPE}
\begin{figure*}
\includegraphics[width=\textwidth]{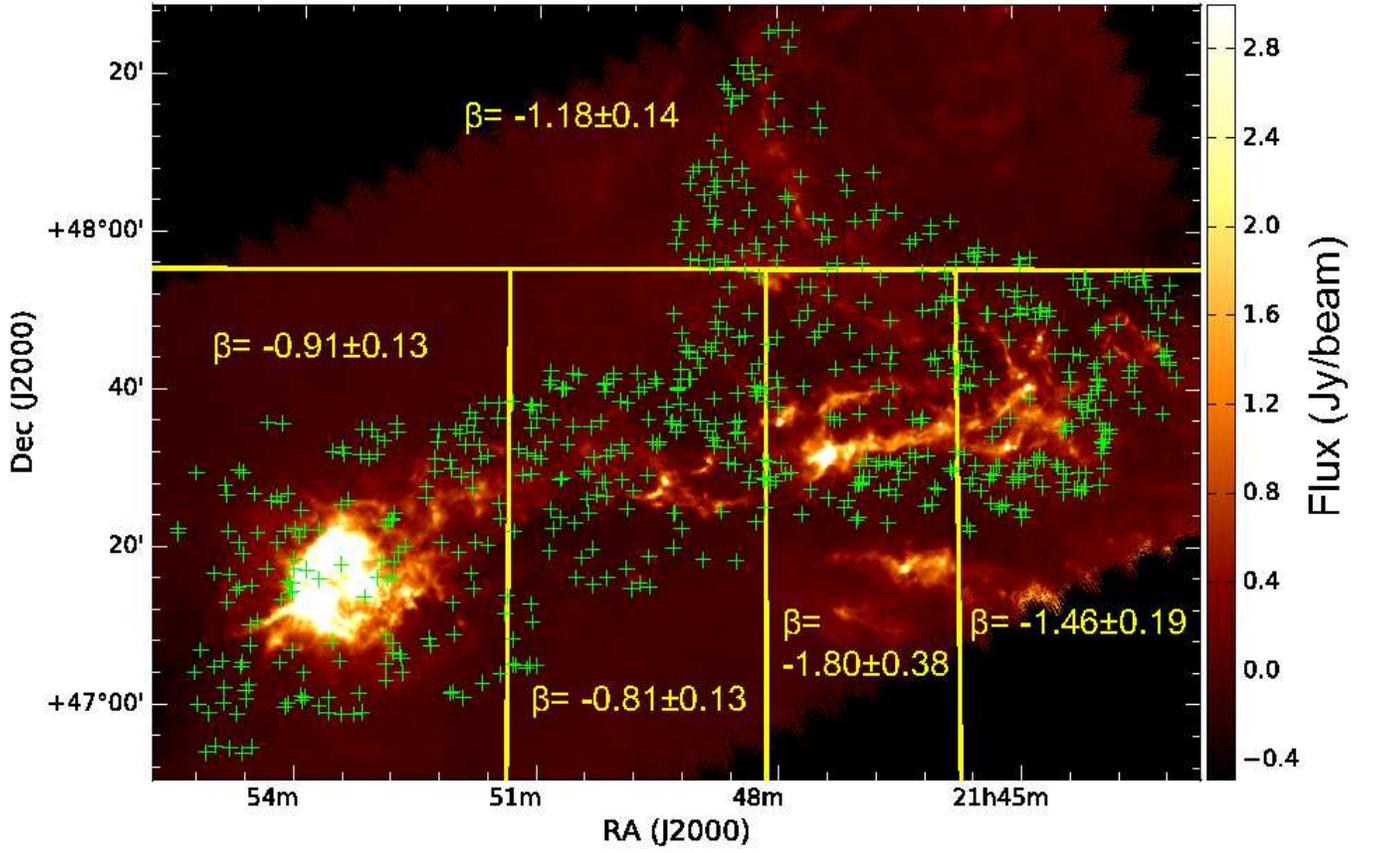}
\caption{PE-$A_V$ relations for the $A_V < 3$ mag stars in particular regions. The green crosses represent the stars with $H$-band polarization detections and $A_V < 3$ mag. These stars were separated into five zones, defined by the yellow lines, and the PE power-law $\beta$ index was calculated for each zone by fitting a single power-law. The $\beta$ values vary with region from -0.81$\pm$0.13 to -1.80$\pm$0.38.}\label{fig:HPeff_diffuse}
\end{figure*}

It is interesting that a variety of PE versus $A_V$ power-law behaviors are seen across the IC5146 cloud system. 
We showed that the indices derived using $H$-band varied with region location in \autoref{fig:HPeff_diffuse}. 
The diverse PE-$A_V$ relations likely depend more on $A_V$ and region choice and less on wavelength, 
as the indices derived from $i'$- and $H$-bands in the same region are almost the same. 
If all the stellar values from all of the low $A_V$ regions are combined into a single plot and fitted, the location dependence of the PE-$A_V$ relation would be mixed and thereby lost. The index of $-0.95$ derived from $H$-band using all data with low $A_V$ is merely the average from the mixed PE-$A_V$ relation shown in \autoref{fig:HPeff_diffuse}. This is different from the suggestion by \citet{go95} that the power-law index of $\sim$ -1 indicates that starlight polarization only traces the dust grains on the surfaces of clouds.

Three possibilities could explain the origin of the wide variation of the PE-$A_V$ relation. 

i) The grain size distributions could be diverse and vary greatly from region to region for low $A_V$. In \autoref{sec:Pspec}, we showed that the dispersion of $\lambda_{max}$, estimated from $P_{R_c}/P_H$, is significantly larger for $A_V < 2.5$ mag, suggesting that grain size distributions are diverse. This diversity is likely region-dependent, since the different $\lambda_{max}$ values were derived from different sightlines. The grain size distribution is a key element of the RATs theory; radiation can efficiently align dust grains having sizes comparable to its wavelength, and wavelength determines radiation penetration ability.

ii) Depolarization effects may occur if multiple polarizing layers exist. In \autoref{fig:HPeff_diffuse}, we showed that the PE vs $A_V$ indices derived in the western part of IC5146 cloud were steeper than $-1$, hence the polarization percentage must decrease with $A_V$. This may be a result of depolarization. If a multiple layer structure, with the layers having different B-field orientations, exists along a line of sight, the net transferred polarization from the different layers would be reduced. \citet{ar11} found rich and hierarchical filamentary structure in the IC5146 cloud system. These filaments can naturally produce a multiple layer structure if the projections of such structures overlap along the line of sight. In addition, they further found that most of the filaments in the western part of the IC5146 cloud system were supercritical, which might twist to cause complex B-fields. This could explain the steeper indices found in the western part of the cloud.

iii) The degree of dust alignment may be affected by one or more bright illuminators, such as the stars in the Cocoon Nebula. \citet{ca14} found a possible dependence of PE-$A_V$ index for regions in the L204 cloud 3 with distance from a nearby illuminator. Brighter stellar radiation can boost alignment of the dust, and the alignment efficiency will mainly depend on the optical depth to the illuminator. In IC5146, the Cocoon Nebula stellar cluster could be a strong illuminator that is able to affect grain alignment. However, the radiation from the Cocoon Nebula is also highly shadowed by the main filament structure, making the optical depth from the Cocoon Nebula to the other regions difficult to estimate.

\subsection{The Breakpoint in PE-$A_V$ Relations}\label{sec:breakpoint}
In \autoref{sec:Peff}, we showed that PE decays with $A_V$ and a breakpoint (BP) of $A_V=2.88\pm0.67$ mag separates two different power-law slopes of $A_V$ in $H$-band. 
Breakpoints of PE near $A_V \sim 2$--3 mag are rarely seen. 
Pipe-109 shows a breakpoint at 9.5 mag with power index changing from -1.00 to -0.34 \citep{al14}. LDN 183 shows a breakpoint at $A_V \sim 20$ mag with power index changing from -0.6 to -1 \citep{cr04,cl12,an15}. \citet{jo15} also found a break point at $A_V \sim 20$ mag with power index changing from -0.5 to -1 toward several starless cores. \citet{jo16} found an break point at very high $A_V$, $\sim$ hundreds mag, in the Class 0 YSO G034.43+00.24 MM1 with a change of power index from -1 to -0.5.

Three mechanisms could change the power-law indices with $A_V$, though in opposite directions.
Radiation into the deepest regions could be too faint to align the dust grains, so the power-law index for high $A_V$ regions should steepen \citep{wh08}.
But, if grains grow at higher $A_V$, the alignment efficiency will be higher for dust with size comparable 
to the wavelength \citep{la07}, hence the power-law index will become flatter in the dense regions where only radiation with longer wavelengths can penetrate and the dust grains are believed to grow. In addition to the change of dust properties, internal radiation fields from embedded sources within dense clouds could also enhance dust alignment, and flatten the power-law index.

\citet{an15} suggest that the origin of the breakpoint found in LDN 183 is likely due to the extinction of external radiation, because their power-law index in low $A_V$ regions, of -0.6, is consistent with prediction of the RATs model assuming constant grain size \citep{wh08}. Similar trends were also found within several starless cores by \citet{jo15}, and they further showed that the RATs theory can explain the break point of $\sim 20$ mag, where the dust optical depth, for the wavelength comparable to the maximum grain size, becomes optically thick. 

Our results showed an opposite trend as compared to the case of LDN183: steeper indices ranging from -0.71 to -1.23 were found in low $A_V$ regions while flatter indices ranging from -0.25 to -0.53, consistent with the \citet{wh08} model, were found in high $A_V$ regions, for all observed wavebands. The results from \citet{al14} toward the Pipe-109 and \citet{jo16} in G034.43+00.24 MM1 show similar trends as ours, but with different breakpoints. The flattened slopes in high $A_V$ regions more likely originate from efficient grain alignment due to either grain growth or internal radiation fields. Because Pipe-109 is a starless object but G034.43+00.24 MM1 is a Class 0 YSO, the change of slope for the former likely results mainly from grain growth while internal radiation might play an important role in the latter. Since the breakpoint of $A_V$ $\sim$2.8 mag shown in our data is consistent with the $A_V$ of the regions where the intrinsic dispersion of $\lambda_{max}$ drops significantly ($A_V$=2.5--4.0 mag), our observed trend is more likely caused by grain growth.

An open question is why the breakpoint we measure is so different from the 9.5 mag value in \citet{al14}, although both breakpoints likely result from grain growth. The difference could originate from two possible reasons. First, \citet{al14} lacked samples in low $A_V$ regions, and thus a broken power-law with breakpoint at $A_V\sim 3$ mag would be difficult to identify. Second, the difference may come from different grain growth conditions. \citet{or09} simulated the evolution of dust grain size, based on grain-grain collisions. They found that dust grains with ice-coatings are more likely to aggregate and grow than are grains without ice-coatings, due to the increased surface stickiness. \citet{chi11} found that $H_2O$-ice in IC5146 only exists in regions exhibiting extinctions exceeding $A_V$ $\sim$ 4 mag, a value quite similar to the breakpoint we found. Thus, H$_2$O-ice mantling could induce a breakpoint via enhanced grain growth. Similarly, \citet{wh01} found the observed $R_{V}$ in Taurus changed from $R_{V}$ $\sim$ 3 to 4 around $A_V=$ 3.2 mag, coincident with the extinction threshold for H$_2$O-ice mantling in Taurus, and suggest that mantle growth is an important process in initial grain growth. We conclude that such mantling also enables greater efficiency of dust grain alignment, permitting magnetic field to be traced deeper into clouds than if such mantling is not present.

\section{CONCLUSIONS}
\label{sec:summary}
We performed optical and infrared polarimetry observations toward background stars seen through the IC5146 dark cloud system using AIMPOL at the ARIES Observatory, TRIPOL at the Lulin Observatory, and Mimir at the Lowell Observatory Perkins Telescope. A total of 2022 stars showed significant polarization detection in at least one of the four wavebands observed. From analysis of these data, we found:

\begin{enumerate}
\item Polarization efficiency (PE $\equiv P_{\lambda}/A_V$) decreases with $A_V$ as a power-law or broken power-law. The values of the power-law indices likely depend more on the choice of targeted regions and local extinction, and less on observing wavelength. 

\item A cloud averaged power-law index of $-0.95$ for PE vs. $A_V$ was found in $H$-band for low $A_V$, the same as the index found in L1755 by \citet{go95}. However, we showed that the index of $-0.95$ resulted from the admixture of a variety of PE-$A_V$ relations, whose local-regional indices varied from -0.8 to -1.8, different from the suggestion by \citet{go95} that the index of $\sim$ -1 indicates that background starlight polarization only traces the dust grains on the surface of clouds.

\item A broken power-law relation for PE vs.\ $A_V$ in $H$-band exhibits a breakpoint at about $A_V$ $\sim$ 2--3 mag. The power-law index in high $A_V$ regions is shallow and consistent with predictions from RATs models \citep{wh08}, while the indices in low $A_V$ regions are steeper and vary with region. The shallow index in high $A_V$ regions is likely due to grain growth.

\item Excess infrared polarization, over that predicted by the Serkowski relation, was observed in $P_{H}/P_K$, possibly resulted from abundance enhancements of large, micron-size dust grains. The average $P_{H}/P_K$ varies with $A_V$ from  
0.96$\pm$0.20 to 1.71$\pm$0.21, and exhibits its highest value of 1.71$\pm$0.21 for $A_V$ in the 2.5--4.0 mag range. This implies that the abundance of such micron-size dust grains might change with $A_V$.

\item Polarization color ($P_{R_c}/P_{H}$) is a useful tool to constrain $\lambda_{max}$, the peak of the Serkowski relation, which may trace the small-size cutoff of the grain size distribution. We found both the average and dispersion of $P_{R_c}/P_{H}$ decreased from the 1.0 $< A_V <$ 2.5 mag range to the 2.5 $< A_V <$ 4.0 mag range. These variations suggest that the small-size cutoffs of grain size distributions are most likely due to efficient grain-grain collisions, and thus hint that grain growth could already take place by $A_V \sim$ 2.5--4.0 mag, possibly enhanced at this relatively low $A_V$ range by the presence of ice mantles on the grains.

\end{enumerate}

In conclusion, this study revealed that dust grains in the diffuse molecular regions ($A_V < 2.5$ mag) of the IC5146 dark cloud system have diverse size distributions. As $A_V$ approaches $\sim 3$ mag, sub-micron dust grains grow significantly due to grain-grain collisions. The size distributions may also become more uniform as $A_V$ increases as a balance is reached between fragmentation and coagulation. In addition, the larger, micron-sized grains likely already exist in the diffuse regions, and their abundances and sizes likely change with $A_V$. In the next paper of this series, we will use these polarization data to characterize the magnetic field morphology and strength across this system of filamentary dark molecular clouds associated with IC5146.

\acknowledgments
We acknowledge support from the Aryabhatta Research Institute of Observational Sciences and Lulin Observatory concerning the data collected using AIMPOL and TRIPOL. This research was conducted in part using the Mimir instrument, jointly developed at Boston University and Lowell Observatory and supported by NASA, NSF, and the W.M. Keck Foundation. This work and the analysis software for Mimir data were developed under NSF grants AST 06-07500, 09-07790, and 14-12269 to Boston University. We thank Brian Taylor, Tao-Chung Ching, and Lauren Cashman for their help in the smooth operations of Mimir observations. This research has made use of the Herchel Science Archive (HSA), and the data are based on the observations performed with the ESA \textit{Herschel} Space Observatory (Pilbratt et al.\ 2010). J.W.W., S.P.L., and C.E. are thankful to the support from the Ministry of Science and Technology (MOST) of Taiwan through the grants NSC 99-2923-M-008-002-MY3, NSC 101-2119-M-007-004, MOST 102-2119-M-007-004-MY3, 105-2119-M-007-022-MY3, and 105-2119-M-007-024. 

\facilities{ARIES:ST, LO:1m, Perkins}
\software{IRAF, Source Extractor \citep{be96}, Mimir Software Package \citep{cl12a}, PNICER \citep{me17}, Astropy \citep{as13}, NumPy \citep{va11}, SciPy \citep{jo01}, Aplpy \citep{ro12}}

\capstartfalse

\begin{table*}[h!]
\centering
\renewcommand{\thetable}{\arabic{table}}
\caption{Measured polarization properties of stars probing the IC5146 dark cloud complex}\label{tab:cat}
\tablenum{1}
\begin{splittabular}{ccccccccccccccccBccccccccccccccccccBccccccccccccc}
\hline
\hline
 & & & \multicolumn{8}{c}{$R_c$-band Polarimetry} & \multicolumn{8}{c}{$i'$-band Polarimetry} & \multicolumn{8}{c}{$H$-band Polarimetry} & \multicolumn{8}{c}{$K$-band Polarimetry} & \multicolumn{6}{c}{2MASS Catalog} &  \multicolumn{4}{c}{WISE Catalog} & & \\
\cmidrule(lr){4-11} \cmidrule(lr){12-19} \cmidrule(lr){20-27} \cmidrule(lr){28-35} \cmidrule(lr){36-41} \cmidrule(lr){42-45}
ID & R.A. & Dec. & $Q_{R_c}$ & $\sigma_{Q_{R_c}}$ & $U_{R_c}$ & $\sigma_{U_{R_c}}$ & $P_{R_c}$ & $\sigma_{P_{R_c}}$ & $PA_{R_c}$ & $\sigma_{PA_{R_c}}$ & $Q_{i'}$ & $\sigma_{Q_{i'}}$ & $U_{i'}$ & $\sigma_{U_{i'}}$ & $P_{i'}$ & $\sigma_{i'}$ & $PA_{i'}$ & $\sigma_{PA_{i'}}$ & $Q_{H}$ & $\sigma_{Q_{H}}$ & $U_{H}$ & $\sigma_{U_{H}}$ & $P_{H}$ & $\sigma_{H}$ & $PA_{H}$ & $\sigma_{PA_{H}}$ & $Q_{K}$ & $\sigma_{Q_{K}}$ & $U_{K}$ & $\sigma_{U_{K}}$ & $P_{K}$ & $\sigma_{P_{K}}$ & $PA_{K}$ & $\sigma_{PA_{K}}$ & J & $\sigma_J$ & H & $\sigma_H$ & K & $\sigma_K$ & W1 & $\sigma_{W1}$ & W2 & $\sigma_{W2}$ & $A_V$\tablenotemark{a} & UF \\
 & (deg) & (deg) & (\%) & (\%) & (\%) & (\%) & (\%)  & (\%)  & (deg)  & (deg) & (\%) & (\%) & (\%) & (\%) & (\%)  & (\%)  & (deg)  & (deg) & (\%) & (\%) & (\%) & (\%) & (\%)  & (\%)  & (deg)  & (deg) & (\%) & (\%) & (\%) & (\%) & (\%)  & (\%)  & (deg)  & (deg) & (mag) & (mag) & (mag) & (mag) & (mag) & (mag) & (mag) & (mag) & (mag) & (mag) & (mag) & \\
(1) & (2) & (3) & (4) & (5) & (6) & (7) & (8) & (9) & (10) & (11) & (12) & (13) & (14) & (15) & (16) & (17) & (18) & (19) & (20) & (21) & (22) & (23) & (24) & (25) & (26) & (27) & (28) & (29) & (30) & (31) & (32) & (33) & (34) & (35) & (36) & (37) & (38) & (39) & (40) & (41) & (42) & (43) & (44) & (45) & (46) & (47) \\
\hline
 8327 & 326.3935 & 47.6311 & 1.68 & 0.32 & 1.94 & 0.33 & 2.54 & 0.34 & 24.51 & 3.44 & 1.59 & 0.07 & 2.18 & 0.07 & 2.70 & 0.07 & 26.90 & 0.70 & 0.41 & 0.11 & 0.65 & 0.11 & 0.76 & 0.11 & 28.62 & 4.28 & 0.03 & 0.22 & 0.53 & 0.20 & 0.49 & 0.20 & 43.34 & 11.76 & 10.861 & 0.019 & 9.727 & 0.028 & 9.352 & 0.017 & 9.163 & 0.023 & 9.314 & 0.019 & 5.00 & 1 \\
12997 & 326.6920 & 47.4848 & 0.98 & 0.24 & 0.90 & 0.24 & 1.31 & 0.25 & 21.34 & 4.86 & 0.54 & 0.05 & 1.40 & 0.05 & 1.50 & 0.05 & 34.46 & 0.97 & 0.47 & 0.23 & 0.47 & 0.22 & 0.62 & 0.23 & 22.28 & 10.49 & 0.31 & 0.25 & 0.44 & 0.16 & 0.51 & 0.20 & 27.36 & 11.20 & 11.483 & 0.025 & 10.745 & 0.03 & 10.462 & 0.023 & 10.257 & 0.022 & 10.35 & 0.021 & 3.03 & 1 \\
13793 & 326.7371 & 47.4965 & 2.11 & 0.18 & 1.40 & 0.18 & 2.53 & 0.18 & 16.75 & 1.97 & 2.16 & 0.05 & 1.75 & 0.05 & 2.78 & 0.05 & 19.55 & 0.50 & 0.98 & 0.21 & 0.70 & 0.23 & 1.19 & 0.22 & 17.82 & 5.33 & 0.37 & 0.24 & 0.36 & 0.13 & 0.48 & 0.19 & 21.98 & 11.66 & 11.205 & 0.023 & 10.333 & 0.028 & 10.075 & 0.022 & 9.894 & 0.022 & 9.961 & 0.021 & 3.16 & 1 \\
\hline
\end{splittabular}
\tablecomments{This table is available in its entirety in a machine-readable form in the online journal. A portion is shown here for guidance regarding its form and content.}
\tablenotetext{a}{The uncertainties of $A_V$ are 0.93 mag, estimated in Section 3.1}
\end{table*}

\capstarttrue
\capstartfalse
\begin{rotate}
\begin{deluxetable}{cccccc}
\tablecaption{Fitting results for PE vs. $A_V$}\label{tab:PE}
\tablenum{2}
\tablehead{\colhead{Band} & \colhead{F-value} & \colhead{p-value} & \colhead{AICc} & \colhead{AICc} & \colhead{Preferred Model} \\ 
\colhead{} & \multicolumn{2}{c}{(model 1 $\rightarrow$ model 2)} & \colhead{(model 1)} & \colhead{(model 2)} & \colhead{} } 
\startdata
$R_c$ & 0.30 & 0.75 & 22.30 & 30.00 & \parbox[t]{7.5cm}{$PE=(0.86\pm0.12)A_V^{(-0.71\pm0.10)}$\\} \\
\hline
i' & 0.40 & 0.68 & -5.64 & 5.74 & \parbox[t]{7.5cm}{$PE=(1.66\pm0.17)A_V^{(-1.23\pm0.10)}$} \\
\hline
H & 3.95 & 0.04 & 40.68 & 39.92 & \parbox[t]{7.5cm}{$PE=\begin{cases}(0.60\pm 0.13) A_V^{(-0.95\pm 0.30)} & : A_V < 2.88\pm0.67\\(0.29\pm 0.07)A_V^{(-0.25\pm 0.06)} & : A_V > 2.88\pm0.67\end{cases}$} \\
\hline
K & 1.13 & 0.36 & 8.90 & 14.57 & \parbox[t]{7.5cm}{$PE=(0.36\pm0.06)A_V^{(-0.53\pm0.09)}$\\} \\
\enddata
\tablecomments{The preferred model is determined if (1) the p value from F-test is below 0.05 (95\% confidence level) or (2) the model has lower AICc value}
\end{deluxetable}
\end{rotate}
\capstarttrue
\capstartfalse
\begin{deluxetable*}{ccccccc}
\tablecaption{Variation of $P_{R_c}/P_{H}$ with $A_V$}\label{tab:pp_RH}
\tablenum{3}
\tablehead{
\colhead{} & \colhead{} & \multicolumn{4}{c}{$P_{R_c}/P_{H}$} & \colhead{} \\
\cline{3-6}
\colhead{Group} & \colhead{$A_V$ range} & \colhead{Mean} & \colhead{STD} & \colhead{$\sigma_{obs}$} & \colhead{Intrinsic} & \colhead{$<\lambda_{max}>$\tablenotemark{a}} \\ 
\colhead{} & \colhead{(mag)} & \colhead{} & \colhead{} & \colhead{} & \colhead{Dispersion\tablenotemark{b}} & \colhead{($\mu$m)} } 

\startdata
A & -2.0--1.0 & $1.54\pm0.09$ & $0.72\pm0.11$ & 0.53 & $0.54\pm0.15$ & $0.92\pm0.02$ \\
B & 1.0--2.5 & $2.69\pm0.20$ & $1.22\pm0.30$ & 0.97 & $0.73\pm0.50$ & $0.75\pm0.03$ \\
C & 2.5--4.0 & $2.48\pm0.12$ & $0.79\pm0.17$ & 0.80 & ... & $0.78\pm0.02$ \\
D & $>$4.0 & $2.81\pm0.22$ & $0.90\pm0.29$ & 0.78 & $0.44\pm0.59$ & $0.73\pm0.03$ \\
\enddata
\tablenotetext{a}{The $\lambda_{max}$ calculated from the mean $P_{R_c}/P_{H}$. Uncertainties were calculated from the means and the propagated uncertainties of the means, assuming c=2.3.}
\tablenotetext{b}{The intrinsic dispersions and the uncertainties of the dispersions. The intrinsic dispersion cannot be defined if the observed dispersion is less than instrumental uncertainty.}
\end{deluxetable*}
\capstarttrue
\capstartfalse
\begin{deluxetable*}{cccccccc}
\tablecaption{Variation of $P_{H}/P_{K}$ with $A_V$}\label{tab:pp_HK}
\tablenum{4}
\tablehead{
\colhead{} & \colhead{} & \multicolumn{4}{c}{$P_{H}/P_{K}$} & \colhead{} & \colhead{} \\
\cline{3-6}
\colhead{Group} & \colhead{$A_V$ range} & \colhead{Mean} & \colhead{STD} & \colhead{$\sigma_{obs}$} & \colhead{Intrinsic} & \colhead{$<\lambda_{max}>$\tablenotemark{a}} & \colhead{Power-law index} \\ 
\colhead{} & \colhead{(mag)} & \colhead{} & \colhead{} & \colhead{} & \colhead{Dispersion\tablenotemark{b}}& \colhead{($\mu$m)} & \colhead{}} 

\startdata
A & -2.0--1.0 & 0.66$\pm$0.12 & 0.39$\pm$0.21 & 0.31 & 0.27$\pm$0.30 & $2.20\substack{+0.13 \\ -0.12}$ & $-1.38\pm0.60$ \\
B & 1.0--2.5 & 0.96$\pm$0.20 & 0.42$\pm$0.31 & 0.52 & ... & $1.93\substack{+0.17 \\ -0.15}$ & $-0.14\pm0.69$ \\
C & 2.5--4.0 & 1.71$\pm$0.21 & 0.57$\pm$0.24 & 0.72 & ... & $1.40\substack{+0.14 \\ -0.13}$ & $1.78\pm0.41$ \\
D & 4.0--10.0 & 1.31$\pm$0.13 & 0.54$\pm$0.20 & 0.54 & ... & $1.67\substack{+0.09 \\ -0.08}$ & $0.90\pm0.33$ \\
E & $>$10.0 & 1.63$\pm$0.16 & 0.51$\pm$0.19 & 0.58 & ... & $1.46\substack{+0.11 \\ -0.11}$ & $1.62\pm0.33$ \\
\enddata
\tablenotetext{a}{The $\lambda_{max}$ calculated from the mean $P_{R_c}/P_{H}$. Uncertainties were calculated from the means and the propagated uncertainties of the means, assuming c=2.3.}
\tablenotetext{b}{The intrinsic dispersions and the uncertainties of the dispersions. The intrinsic dispersion cannot be defined if the observed dispersion is less than instrumental uncertainty.}
\end{deluxetable*}
\capstarttrue

\clearpage


\begin{thebibliography}

\bibitem[Akaike\ (1974)]{ak74}Akaike, H.\ 1974, IEEE Transactions on Automatic Control, 19, 716
\bibitem[Alves et al.\ (2014)]{al14}Alves, F., Frau, P., Girart, J. M., et al.\ 2014, \aap, 569, \bibitem[Andersson \& Potter(2007)]{an07} Andersson, B.-G., \& Potter, S.~B.\ 2007, \apj, 665, 369 
\bibitem[Andersson et al.(2015)]{an15} Andersson, B.-G., Lazarian, A., \& Vaillancourt, J.~E.\ 2015, \araa, 53, 501
\bibitem[Arzoumanian et al.\ (2011)]{ar11}Arzoumanian, D., Andr\'{e}, P., Didelon, P., et al.\ 2011, \aap, 529, L6
\bibitem[Astropy Collaboration et al.(2013)]{as13} Astropy Collaboration, Robitaille, T.~P., Tollerud, E.~J., et al.\ 2013, \aap, 558, A33 
\bibitem[Bertin \& Arnouts (1996)]{be96}Bertin, E., \& Arnouts, S. 1996, \aaps, 117, 393
\bibitem[Cashman \& Clemens (2014)]{ca14}Cashman, L. R., \& Clemens, D. P. 2014, \apj, 793, 126
\bibitem[Chapman et al.\ (2011)]{ch11}Chapman, N. L., Goldsmith, P. F., Pineda, J. L., et al.\ 2011, \apj, 741, 21
\bibitem[Chiar et al.\ (2011)]{chi11}Chiar, J. E., Pendleton, Y. J., Allamandola, L. J., et al.\ 2011, \apj, 731, 9
\bibitem[Cho \& Lazarian(2005)]{ch05} Cho, J., \& Lazarian, A.\ 2005, \apj, 631, 361 
\bibitem[Clayton et al.(1992)]{cl92} Clayton, G.~C., Anderson, C.~M., Magalhaes, A.~M., et al.\ 1992, \apjl, 385, L53 
\bibitem[Clemens et al.\ (2007)]{cl07}Clemens, D. P., Sarcia, D., Grabau, A., et al.\ 2007, PASP, 119, 1385
\bibitem[Clemens(2012)]{cl12} Clemens, D.~P.\ 2012, \apj, 748, 18 
\bibitem[Clemens et al.\ (2012a)]{cl12a}Clemens, D. P., Pinnick, A. F., Pavel, M. D., Taylor, B. W., 2012a, \apjs, 200, 19
\bibitem[Clemens et al.\ (2012b)]{cl12b}Clemens, D. P., Pinnick, A. F., \& Pavel, M. D. 2012b, \apjs, 200, 20
\bibitem[Clemens et al.\ (2012c)]{cl12c} Clemens, D. P., Pavel, M.~D., \& Cashman, L.~R.\ 2012c, \apjs, 200, 21 
\bibitem[Clemens et al.(2016)]{cl16} Clemens, D.~P., Tassis, K., \& Goldsmith, P.~F.\ 2016, \apj, 833, 176 
\bibitem[Crutcher(2004)]{cr04} Crutcher, R.~M.\ 2004, \apss, 292, 225 
\bibitem[Davis \& Greenstein(1951)]{da51} Davis, L., Jr., \& Greenstein, J.~L.\ 1951, \apj, 114, 206
\bibitem[Dolginov \& Mitrofanov (1976)]{do76}Dolginov, A. Z., \& Mitrofanov, I. G. 1976, Ap{\&}SS, 43, 291
\bibitem[Draine \& Weingartner (1996)]{dr96}Draine, B. T., \& Weingartner, J. C. 1996, \apj, 470, 551
\bibitem[Draine \& Weingartner (1997)]{dr97}Draine, B. T., \& Weingartner, J. C. 1997, \apj, 480, 633
\bibitem[Eswaraiah et al.\ (2011)]{es11}Eswaraiah, C., Pandey, A. K., Maheswar, G., et al.\ 2011, \mnras, 411, 1418
\bibitem[Eswaraiah et al.\ (2012)]{es12}Eswaraiah, C., Pandey, A. K., Maheswar, G., et al.\ 2012, \mnras, 419, 2587
\bibitem[Fisher (1925)]{fi25}Fisher R. A., 1925, Nat, 123, 866
\bibitem[Goodman et al.\ (1995)]{go95}Goodman, A. A., Jones, T. J. B., Lada, E. A. , \& Myers, P. C. 1995, \apj, 448, 748
\bibitem[Griffin et al.\ (2010)]{gr10}Griffin, M.J., Abergel, A., Abreu, A., et al.\ 2010, \AA, 518, L3
\bibitem[Hall (1949)]{ha49}Hall, J. S. 1949, Sci, 109, 166
\bibitem[Hatano et al.(2013)]{ha13} Hatano, H., Nishiyama, S., Kurita, M., et al.\ 2013, \aj, 145, 105 
\bibitem[Heiles(2000)]{he00} Heiles, C.\ 2000, \aj, 119, 923 
\bibitem[Herbig \& Dahm(2002)]{he02} Herbig, G.~H., \& Dahm, S.~E.\ 2002, \aj, 123, 304 
\bibitem[Hiltner (1949a)]{hi49}Hiltner, W. A. 1949a, \apj, 109, 471
\bibitem[Hiltner (1949b)]{hi49b}Hiltner, W. A. 1949b, Sci, 109, 165
\bibitem[Hirashita \& Yan(2009)]{hi09} Hirashita, H., \& Yan, H.\ 2009, \mnras, 394, 1061
\bibitem[Hoang \& Lazarian(2009)]{ho09} Hoang, T., \& Lazarian, A.\ 2009, \apj, 695, 1457  
\bibitem[Indebetouw et al.(2005)]{in05} Indebetouw, R., Mathis, J.~S., Babler, B.~L., et al.\ 2005, \apj, 619, 931 
\bibitem[Jones(1990)]{jo90} Jones, T.~J.\ 1990, \aj, 99, 1894 
\bibitem[Jones \& Spitzer(1967)]{jo67} Jones, R.~V., \& Spitzer, L., Jr.\ 1967, \apj, 147, 943 
\bibitem[Jones et al.(1996)]{jo96} Jones, A.~P., Tielens, A.~G.~G.~M., \& Hollenbach, D.~J.\ 1996, \apj, 469, 740 
\bibitem[Jones et al.(2001)]{jo01} Jones, E., Oliphant, E., Peterson, P., et al.\ 2001,  http://www.scipy.org/
\bibitem[Jones et al.(2015)]{jo15} Jones, T.~J., Bagley, M., Krejny, M., Andersson, B.-G., \& Bastien, P.\ 2015, \aj, 149, 31
\bibitem[Jones et al.(2016)]{jo16} Jones, T.~J., Gordon, M., Shenoy, D., et al.\ 2016, \aj, 151, 156 
\bibitem[Kim \& Martin(1995)]{ki95} Kim, S.-H., \& Martin, P.~G.\ 1995, \apj, 444, 293 
\bibitem[Lada et al.\ (1994)]{la94}Lada, C. J., Lada, E. A., Clemens, D. P., \& Bally, J. 1994, \apj, 429, 694
\bibitem[Lada et al.\ (1999)]{la99}Lada, C. J., Alves, J., \& Lada, E. A. 1999, \apj, 512, 250
\bibitem[Lazarian et al.(1997)]{la97} Lazarian, A., Goodman, A.~A., \& Myers, P.~C.\ 1997, \apj, 490, 273 
\bibitem[Lazarian \& Hoang (2007)]{la07}Lazarian, A., \& Hoang, T. 2007, \mnras, 378, 910
\bibitem[Li \& Greenberg(1997)]{li97} Li, A., \& Greenberg, J.~M.\ 1997, \aap, 323, 566 
\bibitem[Li et al.\ (2013)]{li13}Li, H.-b., Fang, M., Henning, T., \& Kainulainen, J. 2013, \mnras, 436, 3707
\bibitem[Lombardi \& Alves(2001)]{lo01} Lombardi, M., \& Alves, J.\ 2001, \aap, 377, 1023  
\bibitem[Ludden et al.(1994)]{lu94} Ludden, Thomas M., Beal, Stuart L., \& Sheiner, Lewis B, C.\ 1994, Journal of Pharmacokinetics and Biopharmaceutics, 22, 5, 431 
\bibitem[Martin(1989)]{ma89} Martin, P.\ 1989, Interstellar Dust, 135, 55 
\bibitem[Martin \& Whittet(1990)]{ma90} Martin, P.~G., \& Whittet, D.~C.~B.\ 1990, \apj, 357, 113 
\bibitem[Martin et al.(1992)]{ma92} Martin, P.~G., Adamson, A.~J., Whittet, D.~C.~B., et al.\ 1992, \apj, 392, 691 
\bibitem[Meingast et al.(2017)]{me17} Meingast, S., Lombardi, M., \& Alves, J.\ 2017, \aap, 601, A137 
\bibitem[Nagata(1990)]{na90} Nagata, T.\ 1990, \apjl, 348, L13 
\bibitem[Ormel et al.(2009)]{or09} Ormel, C.~W., Paszun, D., Dominik, C., \& Tielens, A.~G.~G.~M.\ 2009, \aap, 502, 845 
\bibitem[Parzen (1962)]{pa62} Parzen, E. 1962, Ann. Math. Statist., 33, 1065
\bibitem[Planck Collaboration et al.(2016c)]{pl15} Planck Collaboration, Ade, P.~A.~R., Aghanim, N., et al.\ 2016, \aap, 586, A138 
\bibitem[Purcell (1979)]{pu79} Purcell, E.~M.\ 1979, \apj, 231, 404 
\bibitem[Rautela et al. (2004)]{ra04}Rautela, B. S., Joshi, G. C., \& Pandey, J. C., 2004, Bull. Astron. Soc. India, 32, 159
\bibitem[Robitaille \& Bressert(2012)]{ro12} Robitaille, T., \& Bressert, E.\ 2012, Astrophysics Source Code Library, ascl:1208.017 
\bibitem[Rosenblatt (1956)]{ro56} Rosenblatt, M. 1956, Ann. Math. Statist., 27, 832
\bibitem[Sato et al. (2011)]{sa11}Sato, S., T. Zenno, T. Nagayama, et al., APRIM 2011, NARIT Conference Series, 1, 413
\bibitem[Sato et al. (2017)]{sa17}Sato, S., T. Zenno, T. Nagayama, et al., 2017, submitted to PASJ
\bibitem[Serkowski (1973)]{se73} Serkowski, K.\ 1973, Interstellar Dust and Related Topics, 52, 145 
\bibitem[Serkowski et al.(1975)]{se75} Serkowski, K., Mathewson, D.~S., \& Ford, V.~L.\ 1975, \apj, 196, 261 
\bibitem[Schaefer et al.\ (2007)]{sc07}Schaefer, B., Collett, E., Smyth, R., et al.\ Am. J. Phys. 75, 163 (2007). 
\bibitem[Schmidt et al.\ (1992)]{sc92}Schmidt, G. D., Elston, R., \& Lupie, O. L., 1992, \aj, 104, 1563
\bibitem[Skrutskie et al.(2006)]{sk06} Skrutskie, M.~F., Cutri, R.~M., Stiening, R., et al.\ 2006, \aj, 131, 1163 
\bibitem[Sugitani et al.\ (2011)]{su11}Sugitani, K., Nakamura, F., Watanabe, M., et al.\ 2011, \apj, 734, 63
\bibitem[Sugiura\ (1978)]{su78} Sugiura, N.\ 1978, Comm. Statist., A7, 13
\bibitem[van der Walt et al.(2011)]{va11} van der Walt, S., Colbert, S.~C., \& Varoquaux, G.\ 2011, “The NumPy array: A structure for efficient numerical computation,” Comput. Sci. Eng. 13(2), 22–30
\bibitem[van Leeuwen(2007)]{va07} van Leeuwen, F.\ 2007, \aap, 474, 653
\bibitem[Voshchinnikov et al.(2013)]{vo13} Voshchinnikov, N.~V., Das, H.~K., Yakovlev, I.~S., \& Il'in, V.~B.\ 2013, Astronomy Letters, 39, 421
\bibitem[Voshchinnikov \& Hirashita(2014)]{vo14} Voshchinnikov, N.~V., \& Hirashita, H.\ 2014, \mnras, 445, 301 
\bibitem[Wardle \& Kronberg(1974)]{wa74} Wardle, J.~F.~C., \& Kronberg, P.~P.\ 1974, \apj, 194, 249 
\bibitem[Whittet \& van Breda(1978)]{wh78} Whittet, D.~C.~B., \& van Breda, I.~G.\ 1978, \aap, 66, 57 
\bibitem[Whittet et al.(1992)]{wh92} Whittet, D.~C.~B., Martin, P.~G., Hough, J.~H., et al.\ 1992, \apj, 386, 562 
\bibitem[Whittet et al.(2001)]{wh01} Whittet, D.~C.~B., Gerakines, P.~A., Hough, J.~H., \& Shenoy, S.~S.\ 2001, \apj, 547, 872
\bibitem[Whittet et al.\ (2008)]{wh08}Whittet, D. C. B., Hough, J. H., Lazarian, A., \& Hoang, T. 2008, \apj, 674, 304
\bibitem[Wilking et al.(1982)]{wi82} Wilking, B.~A., Lebofsky, M.~J., \& Rieke, G.~H.\ 1982, \aj, 87, 695 
\bibitem[Wright et al.(2010)]{wr10} Wright, E.~L., Eisenhardt, P.~R.~M., Mainzer, A.~K., et al.\ 2010, \aj, 140, 1868-1881 

\end{thebibliography}
\end{document}